\documentclass[12pt,letterpaper]{article}%
\usepackage{makeidx}
\usepackage{amsmath}
\usepackage{amsfonts}
\usepackage{amssymb}
\usepackage{nopageno}
\usepackage{graphicx}%
\newtheorem{theorem}{Theorem}

\newtheorem{remark}[theorem]{Remark}

\makeatletter
\def\@oddhead{
\vbox{
\hbox to\hsize{\oddmarkA \oddmarkB \hfill \oddmarkC}
}
}
\makeatother
\def\oddmarkA{{\bf }{}}
\def\oddmarkB{}
\def\oddmarkC{\thepage}
\begin{document}

\begin{center}
\textbf{Thermodynamics and quantum cosmology}

\textbf{Continuous topological evolution of topologically coherent defects }

\bigskip

\textbf{R. M. Kiehn\bigskip}

Emeritus Professor of Physics, University of Houston

\ Retired to Mazan, France

\ http://www.cartan.pair.com

\bigskip

Abstract\bigskip
\end{center}

\begin{quotation}
As a point of departure it is suggested that Quantum Cosmology is a
topological concept independent from metrical constraints. \ Methods of
continuous topological evolution and topological thermodynamics are used to
construct a cosmological model of the present universe, using the techniques
based upon Cartan's theory of exterior differential systems. \ Thermodynamic
domains, which are either Open, Closed, Isolated, or in Equilibrium, can be
put into correspondence with topological systems of Pfaff topological
dimension 4, 3, 2 and 1. \ If the environment of the universe is assumed to be
a physical vacuum of Pfaff topological dimension 4, then continuous but
irreversible topological evolution can cause the emergence of topologically
coherent defect structures of Pfaff topological dimension less than 4. \ \ As
galaxies and stars exchange radiation but not matter with the environment,
they are emergent topological defects of Pfaff topological dimension 3 which
are far from equilibrium. \ DeRham topological theory of period integrals over
closed but not exact exterior differential systems leads to the emergence of
quantized, deformable, but topologically coherent, singular macrostates at all
scales. \ The method leads to the conjecture that dark matter and energy is
represented by those thermodynamic topological defect structures of Pfaff
dimension 2 or less.
\end{quotation}

\part{Cosmological Thermodynamics}

\section{Introduction}

Part I of this article suggests that the concepts of Quantum Cosmology should
be addressed in terms of topological concepts rather than metrical geometric
concepts. \ Gravitational metric concepts enter through congruent subsets of a
thermodynamic topology \cite{rmkpv}. \ A primary objective of this article is
to examine the continuous topological\ evolution of various thermodynamic
systems on a cosmological scale without invoking geometric constraints of
metric. \ As a starting point, it is assumed that thermodynamic systems can be
encoded by exterior differential 1-forms on a 4D variety. \ The environment of
the universe will be considered to be a physical vacuum encoded by a
differential 1-form with a Pfaff topological dimension (or class)
\cite{Schouten} equal to 4. \ A thermodynamic system of Pfaff topological
dimension 4 is considered to be an Open thermodynamic system that can exchange
matter and energy with its neighbors. \ It is a nonequilibrium dissipative
system that supports irreversible evolutionary processes.

Emphasis\ in Part I will be placed upon those processes of continuous
topological evolution which cause observable stars and galaxies to emerge as
metastable topological defects of Pfaff topological (not geometric) dimension
3, embedded in the very dilute cosmological, turbulent, nonequilibrium,
environment of Pfaff topological dimension 4. \ The defect structures are
topologically coherent states with properties similar to those found in
"macroscopic" quantum states. \ Various older versions of the basic ideas may
be found at \cite{RMKBilbao}, \cite{RMKVig2003}, \cite{RMKcos004}. \ Extensive
detail can be found in \cite{vol2}. \ \ 

A secondary objective presented in Part II is to update the concept of
topological quantization \cite{rmkperiods}, which can yield macroscopic,
topologically coherent, structures on both cosmological as well as microscopic
scales. \ The details depend upon the existence of closed but not exact
singular p-forms that can be used as integrands in deRham period integrals.

The cosmology constructed herein is based upon continuous topological
evolution of thermodynamic systems. \ When compared to the "bottom up" methods
used to understand the universe in terms of properties deduced from the
microscopic quantum world of Bose-Einstein condensates, super conductors, and
superfluids \cite{volovik}, the cosmology herein, based upon continuous
topological evolution, is a "top down" method. \ Both approaches are similar
in that they utilize the idea that topological defects can support both
microscopic and macroscopic topologically coherent (quantum) states.
\ Thermodynamically,\ the "bottom up" method involves low temperature
equilibrium systems, while the "top down" method is based upon nonequilibrium
thermodynamic systems. \ 

Stars and Galaxies are not equilibrium systems; they are radiating into the
environment. \ They are domains of Pfaff topological dimension 3, while
isolated or equilibrium systems are Pfaff topological dimension 2 or less.
\ Topological defects of Pfaff topological dimension 3 can be far from
equilibrium, and yet can have long metastable, and observable, lifetimes.
\ The thermodynamic method suggests that "dark matter/energy" could have a
mundane explanation in terms of thermodynamic states of Pfaff dimension 2 or
less, representing isolated thermodynamic systems. \ Isolated or equilibrium
thermodynamic domains do not exchange matter or energy with their neighbors,
but could influence gravitational dynamics. \ This topic will be discussed later.

\subsection{Motivation in terms of a Universal van der Waals gas}

As will be demonstrated, an interesting cosmological model for the universe
can be described in terms of a turbulent, dissipative, nonequilibrium, very
dilute, (topologically universal)\footnote{Homeomorphically equivalent}\ van
der Waals gas near its critical point. \ The motivation for treating cosmology
herein from point of view of topological thermodynamics is based upon remarks
made in the Landau-Lifshitz volume on statistical mechanics \cite{LLthermo}.
\ However, the methods used in this article are not statistical, not quantum
mechanical, not metrical, and instead are based on Cartan's methods of
exterior differential forms and their application to the topology of
thermodynamic systems and their continuous topological evolution
\cite{rmkcontopevol}. \ 

Landau and Lifshitz emphasized that real thermodynamic substances, near the
thermodynamic critical point, exhibit (experimentally) extraordinary large
fluctuations of density and entropy. \ In fact, these authors demonstrate that
for an almost perfect gas near the critical point, the correlations of the
fluctuations can be interpreted as a 1/r potential giving a 1/r$^{2}$ force
law of attraction. \ Hence, as a primitive cosmological model, the almost
perfect gas - such as a very dilute van der Waals gas near the critical point
- yields a reason for both the granularity of the night sky and for the
1/r$^{2}$ force law ascribed to gravitational forces between for massive
aggregates. \ The topological thermodynamic methods used in this current
article lead to a similar possibility: the topological defect structures of a
nonequilibrium environment of Pfaff topological dimension 4 can be related to
a topologically universal structure,\ homeomorphic (deformably equivalent) to
a van der Waals gas. \ 

It is assumed that physical thermodynamic systems can be encoded in terms of
an exterior differential 1-form of Action (potentials) on a 4D\ variety of
independent variables. \ A Jacobian matrix can be generated in terms of the
partial derivatives of the coefficient functions that define the 1-form of
Action. \ When expressed in terms of intrinsic variables, known as the
similarity invariants, the Cayley-Hamilton 4 dimensional characteristic
polynomial of the Jacobian matrix generates a universal thermodynamic phase
function. \ Interesting topological defect structures can be put into
correspondence with constraints placed upon the similarity (curvature
symmetry) invariants generated by the Cayley-Hamilton 4 dimensional
characteristic polynomial. \ These constraints define equivalence classes of
topological properties.

The characteristic polynomial of the Jacobian matrix, or Phase function, can
be viewed as representing a family of implicit hypersurfaces in 4D. \ The
hypersurface has an envelope which, when further constrained to be a minimal
hypersurface, is homeomorphic to the Gibbs surface of a van der Waals gas.
\ Another, but different, topological constraint is associated with those
domains for which the determinant of the Jacobian matrix is zero. This
topological constraint on the characteristic polynomial leads to a cubic
factor that mimics the equation of state for a van der Waals gas. \ Hence this
universal topological method for describing a low density turbulent
nonequilibrium media leads to the setting (mentioned above) examined
statistically by Landau and Lifschitz in terms of classical fluctuations about
the critical point. \ 

To repeat, the model presented herein claims that nonequilibrium topological
defects\ in a nonequilibrium 4 dimensional medium represent the stars and
galaxies, which are gravitationally attracted singularities (correlations of
fluctuations of density fluctuations) of a real gas near its critical point.
Note that the Cartan methods do not impose (\textit{a priori)} a constraint of
a metric, connection, or gauge, but do utilize the topological properties
associated with constraints placed on the similarity invariants of the
universal phase function. \ 

Part I of this 2 part article will focus on the topological features of
thermodynamic systems that can be encoded in terms of a 1-form of Action on a
4D variety, and those processes that cause defects to emerge in terms of
continuous topological (not geometrical) evolution. Part II of this 2 part
article will focus on the methods of (macroscopic) topological quantization in
terms of emergent period integrals of closed but not exact p-forms.

What is missing in this approach (based upon the symmetric similarity
invariants)? \ It is the anti-symmetric features of a thermodynamic systems
which lead to torsion, and the source of charge and spin. \ In a recent
development \cite{rmkpv}, a Physical Vacuum was defined in terms of a vector
space of infinitesimal neighborhoods. \ The sole starting point of the theory
resides with the functional format of a matrix Basis Frame of sixteen
functions which can serve as a basis set for the vector space of
infinitesimals. \ A question remains: \newline"Is there any primitive rational
to choose the functional format of the Basis Frame?" \ It will become apparent
from the current presentation that a possible primitive starting point is to
substitute the Jacobian matrix (constructed from the 1-form of Action that
defines a thermodynamic system) as a starting element of an equivalence class
of Basis Frames. \ 

\begin{remark}
The bottom line is: \ The fundamental starting point for an understanding of
cosmology is thermodynamics, not geometry.
\end{remark}

\section{Topological Thermodynamics}

The topological thermodynamic methods \cite{vol1} used herein are based upon
Cartan's theory of exterior differential forms. \ The topological methods
offer an understanding of the cosmos which is considerably different from the
geometric approach assumed by the metrical theory of general relativity. \ The
thermodynamic view assumes that the physical systems to be studied can be
encoded in terms of a 1-form of Action Potentials, $A,$ on a 4 dimensional
variety of ordered independent variables, $\{\xi^{1},\xi^{2},\xi^{3},\xi
^{4}\}.$ \ The variety supports a differential volume element $\Omega_{4}%
=d\xi^{1}\symbol{94}d\xi^{2}\symbol{94}d\xi^{3}\symbol{94}d\xi^{4}. $ \ No
metric, no connection, no constraint of gauge symmetry is imposed upon the 4
dimensional variety. \ \ Topological constraints will be imposed in terms of
exterior differential systems $\cite{Bryant}$

In order to make the equations more suggestive to the reader, the symbolism
for the variety of independent variables will be of the format $\{x,y,z,t\},$
but be aware that no constraints of metric or connection are imposed upon this
variety. \ For instance, it is NOT assumed that the variety is euclidean. \ In
that which follows another useful formalism of independent variables will be
constructed in terms of the ordered set of similarity invariant functions,
which are given the symbols $\{X_{M},Y_{G},Z_{A},T_{K}\}.$ \ The similarity
invariant functions are those deduced from the Jacobian matrix of the
coefficients of that 1-form of Action, $A,$ which is presumed to encode the
thermodynamic properties of a physical system. \ \ 

The 1-form of Action, $A,$ will have components that form a covariant
direction field, $A_{k}(x,y,z,t),$ to within a nonzero factor. \ Evolutionary
processes will be determined in terms of 4 dimensional contravariant direction
fields, $\mathbf{V}_{4}(x,y,z,t)$, to within a nonzero factor. \ Continuous
topological evolution \cite{rmkcontopevol} will be defined in terms of
Cartan's magic formula for the Lie differential, which, when acting on an
exterior differential 1-form of Action, \thinspace$A=A_{k}dx^{k}$, is
equivalent \textit{abstractly} to the first law of thermodynamics. \
\begin{align}
\text{Cartan's Magic Formula \ \ \ }L_{(\mathbf{V}_{4})}A  &  =i(\mathbf{V}%
_{4})dA+d(i(\mathbf{V}_{4})A)\\
\text{First Law of Thermodynamics \ \ \ \ }  &  :W+dU=Q,\\
\text{Inexact 1-form of Heat \ \ \ }L_{(\mathbf{V}_{4})}A  &  =Q\\
\text{Inexact 1-form of Work\ \ \ \ \ \ \ \ \ \ \ \ }W  &  =i(\mathbf{V}%
_{4})dA,\\
\text{Internal Energy \ \ \ \ \ \ \ \ \ \ \ \ \ \ \ \ \ \ }U  &
=i(\mathbf{V}_{4})A.
\end{align}
In effect, Cartan's methods establish a topological basis of thermodynamics in
terms of a theory of cohomology. \ The methods can be used to formulate
precise mathematical definitions for many thermodynamic concepts in terms of
topological properties - without the use of statistics or metric constraints.
\ Moreover, the method applies to nonequilibrium thermodynamical systems and
irreversible processes, again without the use of statistics or metric constraints.

\subsection{The Pfaff Topological Dimension}

One of the most useful topological properties that can be used with exterior
differential forms is that property defined as the Pfaff topological
dimension. \ The Pfaff topological dimension is related to the minimal number
M of functions required\ to define the topological properties of the given
form in a pregeometric variety of dimension N. \ Recall that it is possible to
define many (simultaneous) topologies on the same set of elements. \ For any
(or each) given exterior differential 1-form of functions, say $A=A_{k}%
(x,y,z,t)dx^{k},$ it is possible to construct the Pfaff sequence of terms,
$\{A,dA,A\symbol{94}dA,dA\symbol{94}dA\}.$ \ These elements may be used to
construct a Cartan Topology (relative to the specific 1-form chosen
\cite{baldwin}). \ In the Cartan topology, the exterior differential acts as
limit point generator. \ Hence the union of a form and its exterior
differential create the topological closure of the form \cite{vol1}.

For any given 1-form, the Pfaff sequence will contain $M$ successive nonzero
terms equal to or less than $N$, the number of geometric dimensions of the
base independent variables. \ \ The number $M$ is defined as the "Pfaff
topological dimension", or class, of the given 1-form. \ The three important
1-forms of thermodynamics, $A,\ W,$ and $Q,$ can have different Pfaff
dimensions. \ Suppose the 1-form of work is defined in terms of two functions
as $W=PdV.$ \ The Pfaff sequence consists of the terms $\{W,dW,0,0\};$ hence
in this example, the Pfaff dimension of $W$ is 2. \ From the first law, under
the assumption that $W=PdV,$
\begin{align}
Q  &  =W+dU=PdV+dU,\\
dQ  &  =dW=dP\symbol{94}dV,\\
Q\symbol{94}dQ  &  =W\symbol{94}dW+dU\symbol{94}dW=0+dU\symbol{94}%
dP\symbol{94}dV\\
dQ\symbol{94}dQ  &  =0.
\end{align}
Hence, a Pfaff dimension of 2 for the work 1-form can be associated with a
Pfaff dimension of 3 for the Heat 1-form, unless the Pressure is a function of
the internal energy and the volume. \ In this latter case, the Pfaff dimension
of $Q$ and $W$ are both 2.

In this article, attention will be focused on dissipative turbulent systems
with thermodynamic irreversible processes such that the Pfaff topological
dimensions of $A,\ W,\ $and $Q$ will be maximal and equal to 4. \ (The
techniques can be extended to higher dimensional spaces.) \ These turbulent
systems of Pfaff dimension 4 are not topologically equivalent to Equilibrium
or Isolated systems (for which the topological dimension is 2, at most).
\ Topological defects in the turbulent state will be associated with embedded
sets of space time where the Pfaff topological dimensions are not maximal.
\ It is remarkable that such topological defect sets can form attractors
causing self organization and long lived states of Pfaff dimension 3, which
are far from equilibrium. \ These defects are to be associated with the
emergence of the observable stars and galaxies.

\subsection{Physical Systems}

\subsubsection{Isolated, Closed and Open Systems}

Physical systems and processes are elements of topological categories
determined by the Pfaff topological dimension (or class) of the 1-forms of
Action, $A,$ Work, $W$, and Heat, $Q.$ \ For example, the Pfaff topological
dimension of the exterior differential 1-form of Action, $A,$ determines the
various species of thermodynamic systems in terms of distinct topological
categories: \ %

\begin{align}
Systems  &  :\text{defined by the Pfaff dimension of }A=\rho A^{(0)}%
\nonumber\\
A\symbol{94}dA  &  =0\ \ \ \ \ \ \text{Isolated - Pfaff dimension 2}\\
d(A\symbol{94}dA)  &  =0\ \ \ \ \ \ \text{Closed - Pfaff dimension 3}\\
dA\symbol{94}dA  &  \neq0.\ \ \ \ \text{Open - Pfaff dimension 4.}%
\end{align}
In classical thermodynamics it is often stated that isolated systems do not
permit transport of energy or matter to the environment. \ Closed systems
permit energy (radiation) transport, but not matter transport to the
environment. \ Open systems permit both energy and matter transport to the
environment. \ Note that these topological specifications as given above are
determined entirely from the functional properties of the physical system
encoded as a 1-form of Action, $A$. \ The system topological categories do not
involve a process, which is assumed to be encoded by some vector direction
field, $\mathbf{V}_{4}.$ $\ $The cosmological model presented herein is based
on an open, Pfaff dimension 4, nonequilibrium, turbulent physical system, with
internal defect structures of lesser Pfaff topological dimension acting as
stars and galactic mass aggregates\footnote{Could it be that "dark matter" is
simply related to those thermodynamic states which are isolated, and are of
Pfaff topological dimension 2 or less?}.

\subsubsection{Equilibrium vs. Non-Equilibrium Systems}

The intuitive idea for an equilibrium system comes from the experimental
recognition that the intensive variables of pressure and temperature become
domain constants in an equilibrium state: \ $dP\Rightarrow0,\ dT\Rightarrow0$.
\ A definition made herein is that the Pfaff dimension of a physical system in
the equilibrium state is at most 2 \cite{Bamberg}. \ The Cartan topology
generated by the elements of the Pfaff sequence for $A$\ is then a connected
topology of one component, $\{A\neq0,dA\neq0,A\symbol{94}dA=0\}.$ \ Although
the Pfaff dimension of $A$ is at most 2 in the equilibrium state, processes in
the equilibrium state are such that the Work 1-form and the Heat 1-form must
be of Pfaff dimension 1. \ For suppose $W=PdV,$ then $dW=dP\symbol{94}%
dV\Rightarrow0$ if the pressure is a domain constant. \ Similarly, suppose
$Q=TdS,$ the $dQ=dT\symbol{94}dS\Rightarrow0$ if the temperature is a domain
constant. \ Hence both $W$ and $Q$ are of Pfaff dimension 1 for this example.

A more stringent sufficient condition for equilibrium can be constructed in
terms of the structure of the system, valid for any choice of process. \ For
if the Pfaff dimension of the 1-form of Action is 1, then $dA\Rightarrow0.$
\ It follows that $W\Rightarrow0,$ hence the Pressure must vanish, and Heat
1-form is a perfect differential, $Q=d(U).$ \ 

The cosmological model proposed herein presumes that the physical vacuum is of
Pfaff dimension 4, containing defect structures of Pfaff dimension 3, or less.
\ Both non-equilibrium domains of Pfaff dimension 4 or 3 can admit processes
that are thermodynamically irreversible. \ Extremal processes of the
Hamiltonian type do not exist in domains of Pfaff topological dimension 4.
\ The theory of continuous topological evolution indicates that embedded non
equilibrium (they are radiating) topological defects of Pfaff dimension 3, can
emerge in domains of Pfaff topological dimension 4 via irreversible processes.
\ Once formed and self-organized as coherent topological attractors, the
defect structures of Pfaff topological dimension 3 can continue to evolve
along extremal trajectories that are not irreversibly dissipative. \ They can
have finite lifetimes modified by topological fluctuations. \ In this sense,
these topologically coherent defect structures are analogues of "stationary
excited states" far from equilibrium. \ 

The descriptive words of self-organized states far from equilibrium are
abstracted from the intuition and conjectures of I. Prigogine \cite{Prig}.
\ However, the topological theory presented herein presents for the first time
a solid formal justification (with examples) for the Prigogine conjectures.
\ Precise definitions of equilibrium and nonequilibrium systems, as well as
reversible and irreversible processes can be made in terms of the topological
features of Cartan's exterior calculus. Thermodynamic irreversibility and the
arrow of time are well defined in a topological sense \cite{rmkarw}, a
technique that goes beyond (and without)\ statistical analysis.

\subsubsection{Multiple Components}

One of the most remarkable properties of the Cartan topology generated by a
Pfaff sequence is associated with the fact that when $A\symbol{94}dA=0,$
(Pfaff dimension 2 or less) the physical system is reducible to a single
connected topological component. \ On the other hand when $A\symbol{94}%
dA\neq0,$ (Pfaff dimension 3 or more) the physical system admits more than one
topological component. \ The bottom line is that when the Pfaff dimension is 3
or greater (such that conditions of the Frobenius unique integrability theorem
are not satisfied), solution uniqueness to the Pfaffian differential equation,
$A=0,$ is lost. \ If there exist solutions, there is more than one. \ Such
concepts lead to propagating discontinuities (signals), envelope solutions
(Huygen wavelets), an edge of regression and lack of time reversal invariance,
and the existence of irreducible affine torsion in the theory of connections.
\ \ \ It is the opinion of this author that a dogmatic insistence on
uniqueness historically has hindered the understanding of irreversibility and
nonequilibrium systems.

\subsection{Processes}

\subsubsection{Reversible and Irreversible Processes}

The Pfaff topological dimension of the exterior differential 1-form of Heat,
$Q,$ determines important topological categories of processes. \ From
classical thermodynamics "The quantity of heat in a reversible process always
has an integrating factor" \cite{Goldenblat} \cite{Morse} . \ Hence, from the
Frobenius unique integrability theorem, all reversible processes are such that
the Pfaff dimension of $Q$ is less than or equal to 2. \ Irreversible
processes are such that the Pfaff dimension of $Q$ is greater than 2. \ A
dissipative irreversible topologically \textit{turbulent} process is defined
when the Pfaff dimension of $Q$ is 4. \ \
\begin{align}
Processes  &  :\text{defined by the Pfaff dimension }Q\nonumber\\
Q\symbol{94}dQ  &  =0\ \ \ \ \ \text{\ \ Reversible - Pfaff dimension 2 }\\
d(Q\symbol{94}dQ)  &  \neq0.\ \ \ \ \text{\ Turbulent - Pfaff dimension 4. }%
\end{align}
Note that the Pfaff dimension of $Q$ depends on both the choice of a process,
$\mathbf{V}_{4},$ and the system, $A$, upon which it acts. \ As reversible
thermodynamic processes are such that $Q\symbol{94}dQ=0,$ and irreversible
thermodynamic processes are such that $Q\symbol{94}dQ\neq0,$ Cartan's formula
of continuous topological evolution can be used to determine if a given
process, $\mathbf{V}_{4},$ acting on a physical system, $A$, is
thermodynamically reversible or not: \ \
\begin{equation}
\left[
\begin{array}
[c]{c}%
\text{Reversible Processes\ }\mathbf{V}_{4}:\text{\ }L_{(\mathbf{V}_{4}%
)}A\symbol{94}L_{(\mathbf{V}_{4})}dA=0,\\
\text{Irreversible Processes }\mathbf{V}_{4}:L_{(\mathbf{V}_{4})}%
A\symbol{94}L_{(\mathbf{V}_{4})}dA\neq0.
\end{array}
\right]
\end{equation}

In this article it is assumed that the cosmological background for space-time
belongs to the dissipative irreversible turbulent\ nonequilibrium category,
where the Pfaff topological dimension (or class) is maximal and equal to 4,
almost everywhere, for each of the 1-forms of Action, $A,$ Work, $W$, and
Heat, $Q.$ \ Of particular interest will be those subsets of space and time
where the turbulent nonequilibrium category admits, or evolves into,
topological defects such that the Pfaff topological dimension for all three
1-forms is no longer maximal and equal to 4. \ Remarkably, Cartan's magic
formula can be used to describe the continuous dynamic possibilities of both
reversible and irreversible processes, in equilibrium or nonequilibrium
systems, even when the evolution induces topological change, transitions
between excited states, and changes of phase, such as condensations.

It is important to note that the velocity field need not be topologically
constrained such that it is singularly parameterized. \ That is, the
evolutionary processes described by Cartan's magic formula are not necessarily
restricted to vector fields that satisfy the topological constraints of
kinematic perfection, $dx^{k}-V^{k}dt=0$. \ A discussion of topological
fluctuations and an example fluctuation process is described in section 6.

\subsubsection{Adiabatic Processes - Reversible and Irreversible}

The topological formulation permits a precise definition to be made for both
reversible and an irreversible adiabatic processes in terms of the topological
properties of $Q.$ \ On a geometrical space of N dimensions, a 1-form will
admit N-1 vector fields such that $i(V_{A})Q=0.$ \ Such processes $V_{A}$ are
defined as adiabatic processes \cite{Bamberg}. \ Note that adiabatic processes
are defined by vector direction fields, to within an arbitrary factor,
$\beta(x,y,z,t)$. \ That is, if $i(V_{A})Q=0,$ then it is also true that
$i(\beta V_{A})Q=0.$ The differences between the inexact 1-forms of Work and
Heat become obvious in terms of the topological format. \ Both 1-forms depend
on the process and on the physical system. \ However, Work is\ always
transversal to the process, as $i(\mathbf{V}_{4})W=i(\mathbf{V}_{4}%
)i(\mathbf{V}_{4})dA=0,$ but Heat is\ not, as $i(\mathbf{V}_{4})Q=i(\mathbf{V}%
_{4})dU\Rightarrow0,$ only for adiabatic processes.

It is not obvious that the adiabatic direction fields are such that the Pfaff
dimension of $Q$ is 2. \ That is, it is not obvious that $Q$ can be written in
the form, $Q=TdS,$ as is possible on the manifold of equilibrium states.\ From
the Cartan formulation it is apparent that if $Q$ is not zero, then%

\begin{align}
i(\mathbf{V}_{A})L_{(\mathbf{V}_{A})}A  &  =i(\mathbf{V}_{A})i(\mathbf{V}%
_{A})dA+i(\mathbf{V}_{A})d(i(\mathbf{V}_{A})A)\\
&  =0+i(\mathbf{V}_{A})d(i(\mathbf{V}_{A})A)=i(\mathbf{V}_{A})Q
\end{align}
Hence, for an Adiabatic process:%

\begin{equation}
\text{Adiabatic process }0+i(\mathbf{V}_{A})d(i(\mathbf{V}_{A})A)=i(\mathbf{V}%
_{A})Q\Rightarrow0,\ \ Q\neq0.
\end{equation}
A reversible process is defined such that $Q$ is less than Pfaff dimension 3,
or $Q\symbol{94}dQ=0$ \ Hence $i(\mathbf{V}_{A})(Q\symbol{94}dQ)=0.$ \ \ But
\begin{equation}
i(\mathbf{V}_{A})(Q\symbol{94}dQ)=(i(\mathbf{V}_{A})Q)\symbol{94}%
dQ-Q\symbol{94}i(\mathbf{V}_{A})dQ
\end{equation}
which permits reversible and irreversible adiabatic processes to be well
defined\footnote{It is apparent that i(V)Q= 0 defines an adiabatic process,
but not necessarily a reversible adiabatic process. \ This topological point
clears up certain misconceptions that appear in the literature.} when $Q\neq0$:%

\begin{align}
\text{Reversible Adiabatic Process }  &  =-Q\symbol{94}i(\mathbf{V}%
_{A})dQ\Rightarrow0,~i(\mathbf{V}_{A})Q\Rightarrow0,\\
\text{Irreversible Adiabatic Process }  &  =-Q\symbol{94}i(\mathbf{V}%
_{A})dQ\neq0,~\ ~i(\mathbf{V}_{A})Q\Rightarrow0.
\end{align}

It is certainly true that if $L_{(\mathbf{V})}A=Q=0,$ \textit{identically},
then all such processes are adiabatic, and reversible. \ In such cases, the
Cartan formalism implies that $W+dU=0.$ \ \ Such systems are elements of the
Hamiltonian class of processes, where $W=d\Theta$. \ Recall that \textit{all}
Hamiltonian processes are thermodynamically reversible. \ Hamiltonian
processes are adiabatic when the internal energy $U=(i(V)A)$ is an
evolutionary invariant. \ %

\begin{align}
\text{Hamiltonian Adiabatic Process }  &  =L_{(\mathbf{V})}%
\{i(V)A\}=i(V)Q=0,\text{ }\\
W  &  =i(V)dA=d\Theta,\ \ \\
i(V)W  &  =0,\ \ \ \ \ i(V)A=U.
\end{align}

Note that for a given 1-form of heat, $Q$, it is possible to construct a
matrix of N-1 null vectors, and then to compute the adjoint matrix of
cofactors transposed to create the unique direction field (to within a
factor), $\mathbf{V}_{NullAdjoint}.$ \ Evolution in the direction of
$\mathbf{V}_{NullAdjoint}$ does not represent an adiabatic process path, as
$i(\mathbf{V}_{NullAdjoint})Q\neq0$. \ For a given $Q$, the N-1 null vectors
need not span a smooth hypersurface whose surface normal is proportional to a
gradient field. \ The components of the 1-form may be viewed as the normal
vector to an implicit hypersurface, but the implicit hypersurface is not
necessarily defined as the zero set of some function. \ 

\subsubsection{Topological Torsion}

For maximal, nonequilibrium, turbulent systems in space-time, the maximal
element in the Pfaff sequence generated by $A,W,$ or $Q,$ is a 4-form. \ On
the geometric space of 4 independent variables, every 4-form is globally
closed, in the sense that its exterior differential vanishes everywhere. \ It
follows that every 4-form is exact and can be generated by the exterior
differential of a 3-form. \ The exterior differential of the 3-form is related
to the concept of a divergence of a contravariant direction (vector) field.
\ Most of the development in this article will be devoted to the study of such
3-forms, and their kernels. \ It is a remarkable fact that all 3-forms admit
integrating denominators, such that their exterior differential of a rescaled
3-form is zero almost everywhere. \ Space time points upon which the
denominator has a zero value form defect topological structures.

When the Action for a physical system is of Pfaff dimension 4, there exists a
unique direction field, $\mathbf{T}_{4},$ defined as the topological torsion
4-vector, that can be evaluated \textit{entirely} in terms of those component
functions of the 1-form of Action which define the physical system. \ To
within a factor, this direction field\footnote{A direction field is defined by
the components of a vector field which establish the "line of action" of the
vector in a projective sense. \ An arbitrary factor times the direction field
defines the same projective line of action, just reparameterized. \ In metric
based situations, the arbitrary factor can be interpreted as a renormalization
or conformal factor.} has the four components of the 3-form $A\symbol{94}dA,$
with the properties such that%

\begin{align}
i(\mathbf{T}_{4})\Omega_{4}  &  =A\symbol{94}dA\\
W  &  =i(\mathbf{T}_{4})dA=\sigma\ A,\\
U  &  =i(\mathbf{T}_{4})A=0,\\
Q\symbol{94}dQ  &  =L_{(\mathbf{T}_{4})}A\symbol{94}L_{(\mathbf{T}_{4}%
)}dA=\sigma^{2}A\symbol{94}dA\\
dA\symbol{94}dA  &  =2\ \sigma\ \Omega_{4}.
\end{align}
Hence, evolution in the direction of $\mathbf{T}_{4}$ is thermodynamically
irreversible, when $\sigma\neq0$ and $A$ is of Pfaff dimension 4. \ The kernel
of this vector field is defined as the zero set under the mapping induced by
exterior differentiation. \ In engineering language, the kernel of this vector
field are those point sets upon which the divergence of the vector field
vanishes. \ The Pfaff dimension of the Action 1-form is 3 in the defect
regions defined by the kernel of $\mathbf{T}_{4}$.

For purposes of more rapid comprehension, consider a 1-form of Action, $A$,
with an exterior differential, $dA,$ and a notation that admits an
electromagnetic interpretation ($\mathbf{E=-\partial A/\partial}t-\nabla\phi,
$ and $\mathbf{B}=\nabla\times\mathbf{A)}\footnote{The bold letter
$\mathbf{A}$ represents the first 3 components of the 4 vector of potentials,
with the order in agreement with the ordering of the independent variables.
\ The letter A represents the 1-form of Action.}.$ \ The explicit format of
$\ \textbf{T}_4$ becomes:
\begin{align}
\mathbf{T}_{4}  &  =-[\mathbf{E}\times\mathbf{A}+\mathbf{B}\phi,\ \mathbf{A}%
\circ\mathbf{B]}\text{ \ Topological Torsion 4 vector}\mathbf{,}\\
A\symbol{94}dA  &  =i(\mathbf{T}_{4})\Omega_{4}\\
&  ={\small T}_{4}^{x}{\small dy\symbol{94}dz\symbol{94}dt-T}_{4}%
^{y}{\small dx\symbol{94}dz\symbol{94}dt+T}_{4}^{z}{\small dx\symbol{94}%
dy\symbol{94}dt-T}_{4}^{t}{\small dx\symbol{94}dy\symbol{94}dz,}\\
dA\symbol{94}dA  &  =2(\mathbf{E}\circ\mathbf{B)}\ \Omega_{4}\\
&  =\{{\small \partial T}_{4}^{x}{\small /\partial x+\partial T_{4}%
^{y}/\partial y+\partial T_{4}^{z}/\partial z+\partial T_{4}^{t}/\partial
t\}}\ \Omega_{4}\mathbf{.}%
\end{align}
When the divergence of the topological torsion vector is not zero, $%
\sigma=(\textbf{E}\circ\textbf{B)}\neq0,$ and $A$ is of Pfaff dimension 4,
\thinspace$W$ is of Pfaff dimension 4, and $Q$ is of Pfaff dimension 4. \ The
process generated by $\textbf{T}_4$ is thermodynamically irreversible. \ The
evolution of the volume element relative to the irreversible process
$\textbf{T}_4$\ is given by the expression,%

\begin{align}
L(\mathbf{T}_{4})\Omega_{4}  &  =i(\mathbf{T}_{4})d\Omega_{4}+d(i(\mathbf{T}%
_{4})\Omega_{4})\\
&  =0+d(A\symbol{94}dA)=2(\mathbf{E}\circ\mathbf{B)}\ \Omega_{4}.
\end{align}
Hence, the differential volume element (and therefore the turbulent
cosmological universe)\ is expanding or contracting depending on the sign and
magnitude of $\mathbf{E}\circ\mathbf{B.}$ \ In a fluid model, the coefficient
$\mathbf{E}\circ\mathbf{B}$ plays the role of a bulk viscosity coefficient.

If $A$ is (or becomes) of Pfaff dimension 3, then $dA\symbol{94}%
dA\Rightarrow0$ which implies that $\sigma^{2}\Rightarrow0,$ but
$A\symbol{94}dA\neq0.$ \ The differential volume element $\ \Omega_{4}~$\ is
subsequently an evolutionary invariant, and evolution in the direction of the
topological torsion vector is thermodynamically reversible. \ The physical
system is not in equilibrium, but the divergence free $\mathbf{T}_{4}$
evolutionary process forces the Pfaff dimension of $W$ to be zero, and the
Pfaff dimension of $Q$ to be at most 1. \ Indeed, a divergence free
$\mathbf{T}_{4}$ evolutionary process has a Hamiltonian representation. \ \ In
the domain of Pfaff dimension 3 for the Action, $A$, the subsequent continuous
evolution of the system, $A$, relative to the process $\mathbf{T}_{4},$
proceeds in an energy conserving manner, representing a "stationary" or
"excited" state far from equilibrium. \ These excited states can be
interpreted as the evolutionary topological defects in the turbulent
dissipative system of Pfaff dimension 4.

On a geometric domain of 4 dimensions, assume that the evolutionary process
generated by $\mathbf{T}_{4}$ starts from an initial condition (or state)
where the Pfaff topological dimension of $A$ is also 4. \ Depending on the
sign of the divergence of $\mathbf{T}_{4},$ the process follows an
irreversible path for which the divergence represents an expansion or a
contraction. \ If the irreversible evolutionary path is attracted to a region
(or state) where the Pfaff topological dimension of the 1-form of Action is 3,
then $\mathbf{E\circ B}$ becomes (or has decayed to) zero. \ The zero set of
the function $\mathbf{E\circ B}$ defines a hypersurface in the 4 dimensional
space. \ If the process remains trapped on this hypersurface of Pfaff
dimension 3, $\mathbf{E\circ B}$ remains zero, and the $\mathbf{T}_{4}$
process becomes an extremal field. \ Such extremal fields are such that the
virtual work 1-form vanishes, $W=$ $i(\mathbf{T}_{4})dA=0,$ and the now
reversible $\mathbf{T}_{4}$ process has a Hamiltonian representation. \ The
system is conservative in a Hamiltonian sense, but it is in an "excited" state
on the hypersurface that is far from equilibrium, as the Pfaff dimension of
the 1-form of Action is 3, and not 2. \ (Further evolution could lead to limit cycles.)

The fundamental claim made in this article is that it is these topological
defects, that self organize (emerge) from the dissipative irreversible
evolution of the turbulent state into "stationary metastable" states far from
equilibrium, that form the stars and the galaxies of the cosmos. \ They are
the long lived remnants or "wakes" generated from irreversible processes in
the dissipative nonequilibrium turbulent medium.

\section{Thermodynamic Cosmology}

\subsection{The Jacobian Matrix of the Action 1-form.}

The idea is to express the Jacobian matrix of the coefficient functions that
define the 1-form of Action, $A$, in terms of "universal" coordinates.
\ \ These universal coordinates will be the similarity invariants of the
Jacobian matrix. \ For a 1-form of Action of Pfaff topological dimension 4,
the Cayley-Hamilton theorem produces a Universal Phase function as a
polynomial of 4th degree. \ What is remarkable about this Universal Phase
function is that it has properties that are homeomorphically deformable into
the format of a classic van der Waals gas. \ It is this universality that
gives credence to the idea that the universe could be a nonequilibrium van der
Waals gas near its critical point.

\subsubsection{The Universal Characteristic Phase Function}

The 1-form of Action, used to encode a physical system, contains other useful
topological information, as well as geometric information. \ Consider the
turbulent thermodynamic state generated by a 1-form of Action, $A,$ of Pfaff
topological dimension 4. \ \ The component functions of the Action 1-form can
be used to construct a 4x4 Jacobian matrix of partial derivatives, $\left[
\mathbb{J}_{jk}\right]  =[\partial(A_{j})/\partial x^{k}] $. \ In general,
this Jacobian matrix will be a 4 x 4 matrix that satisfies a 4th order
Cayley-Hamilton characteristic polynomial equation, $\Theta(x,y,z,t;\ \Psi
)=0,$ where $\Psi$ is a possibly complex order parameter with 4 perhaps
complex roots $\rho_{k}$ representing the complex eigenvalues of the Jacobian
matrix. \ %

\begin{equation}
\Theta(x,y,z,t;\ \Psi)=\Psi^{4}-X_{M}\Psi^{3}+Y_{G}\Psi^{2}-Z_{A}\Psi
^{1}+T_{K}\Rightarrow0.
\end{equation}
The functions $X_{M}(x,y,z,t),Y_{G}(x,y,z,t),Z_{A}(x,y,z,t),T_{K}(x,y,z,t)$
are the similarity invariants of the Jacobian matrix. \ If the eigenvalues are
distinct, then the similarity invariants are given by the expressions:%
\begin{align}
X_{M}  &  =\rho_{1}+\rho_{2}+\rho_{3}+\rho_{4},\\
Y_{G}  &  =\rho_{1}\rho_{2}+\rho_{2}\rho_{3}+\rho_{3}\rho_{1}+\rho_{4}\rho
_{1}+\rho_{4}\rho_{2}+\rho_{4}\rho_{3},\\
Z_{A}  &  =\rho_{1}\rho_{2}\rho_{3}+\rho_{4}\rho_{1}\rho_{2}+\rho_{4}\rho
_{2}\rho_{3}+\rho_{4}\rho_{3}\rho_{1},\\
T_{K}  &  =\rho_{1}\rho_{2}\rho_{3}\rho_{4}.
\end{align}

The similarity invariants may be considered as a coordinate map from the
original variety of independent variables, $\{x,y,z,t\}\Rightarrow
\{X_{M},Y_{G},Z_{A},T_{K}\}.$ When the similarity invariants are treated as
generalized coordinates, then the characteristic polynomial becomes a
Universal Phase function, and will be used to encode universal thermodynamic properties.

\subsubsection{Minimal surfaces}

The Universal Phase function, $\Theta,$ may be considered as a family of
hypersurfaces in the 4 dimensional space, $\{X_{M},Y_{G},Z_{A},T_{K}\}$ with a
complex family (order) parameter, $\Psi.$ \ Moreover, it should be realized
that the Universal Phase Function is a holomorphic function, $\Theta
=\phi+i\chi$ in the complex variable $\Psi=u+iv.$ \ That is%

\begin{equation}
\Theta(X_{M},Y_{G},Z_{A},T_{K};\ \Psi)\Rightarrow\phi+i\chi,\
\end{equation}
where
\begin{align}
\phi &  =u^{4}-6u^{2}v^{2}+v^{4}-X_{M}(u^{3}-3uv^{2})+Y_{G}(u^{2}-v^{2}%
)-Z_{A}u+T_{K}\\
\chi &  =4u^{3}v-4uv^{3}-X_{M}(3u^{2}v-v^{3})+2Y_{G}uv-Z_{A}v.
\end{align}
As such, in the 4D space of two complex variables, \{$\phi+i\chi,u+iv\},$
according to the theorem of Sophus Lie, any such holomorphic function produces
a pair of conjugate \textit{minimal} surfaces in the 4 dimensional space
$\{\phi,\chi,u,v\}.$ \ It follows that there exist a sequence of maps,
\begin{equation}
\{x,y,z,t\}\Rightarrow\{X_{M},Y_{G},Z_{A},T_{K}\}\Rightarrow\{\phi,\chi,u,v\}
\end{equation}
such that the family of hypersurfaces can be decomposed into a pair of
conjugate minimal surface components. \ The criteria for a minimal surface is
equivalent to the idea that $X_{M}=0.$ \ By suitable renormalization, the
similarity invariant $X_{M}$ is equivalent to the Mean Curvature of the hypersurface.

\subsection{Envelopes}

The theory of implicit hypersurfaces focuses attention upon the possibility
that the Universal Phase function has an envelope. \ The existence of an
envelope depends upon the possibility of finding a simultaneous solution to
the two implicit surface equations of the family:
\begin{equation}
\Theta(x,y,z,t;\ \Psi)=\Psi^{4}-X_{M}\Psi^{3}+Y_{G}\Psi^{2}-Z_{A}\Psi
+T_{K}\Rightarrow0.
\end{equation}%
\begin{equation}
\text{ \ }\partial\Theta/\partial\Psi=\Theta_{\Psi}=4\Psi^{3}-3X_{M}\Psi
^{2}+2Y_{G}\Psi-Z_{A}\Rightarrow0.
\end{equation}
For the envelope to be smooth, it must be true that $\partial^{2}%
\Theta/\partial\Psi^{2}=\Theta_{\Psi\Psi}\neq0,$ and that the exterior 2-form,
$d\Theta\symbol{94}d\Theta_{\Psi}\neq0$ subject to the constraint that the
family parameter is a constant: $d\Psi=0.$ \ The envelope as a smooth
hypersurface does not exist unless both conditions are satisfied. \ \ Recall
that the envelope, if it exists, is a hypersurface in the space of similarity
coordinates, $\{X_{M},Y_{G},Z_{A},T_{K}\}.$

The envelope is determined by the discriminant of the Phase Function
polynomial, which as a zero set is equal to a universal hypersurface in the 4
dimensional space of similarity variables $\{X_{M},Y_{G},Z_{A},T_{K}\}.$
\ This function can be written in terms of the similarity "coordinates"
(suppressing the subscripts) :%

\begin{align}
&  \text{Discriminant of the Universal Phase Function}\nonumber\\
&  {\small =}{\small 18X}^{3}{\small ZYT-27Z}^{4}{\small +Y}^{2}{\small X}%
^{2}{\small Z}^{2}{\small -4Y}^{3}{\small X}^{2}{\small T+144YX}^{2}%
{\small T}^{2}\\
&  {\small +18XZ}^{3}{\small Y-192XZT}^{2}{\small -6X}^{2}{\small Z}%
^{2}{\small T+144TZ}^{2}{\small Y-4X}^{3}{\small Z}^{3}\\
&  {\small -27X}^{4}{\small T}^{2}{\small -4Y}^{3}{\small Z}^{2}%
{\small +16Y}^{4}{\small T-128Y}^{2}{\small T}^{2}{\small +256T}%
^{3}{\small -80XZY}^{2}{\small T.}%
\end{align}

The discriminant has eliminated the family order parameter. \ \ Remarkably, if
the linear similarity invariant related to the Mean Curvature is set to zero,
$X_{M}\Rightarrow0,$ then the constrained discriminant describes a universal
swallow tail surface homeomorphic (deformable) to the Gibbs surface (see the
figure below) of a van der Waals gas (subscripts suppressed):%

\begin{align}
&  \text{Universal Gibbs Swallowtail Envelope }(X=0,Y,Z,T)\\
&  ={\small -27Z}^{4}{\small +144TZ}^{2}{\small Y-4Y}^{3}{\small Z}%
^{2}{\small +16Y}^{4}{\small T-128Y}^{2}{\small T}^{2}{\small +256T}%
^{3}\Rightarrow0.
\end{align}

In other words, the Gibbs function for a van der Waals gas is a universal idea
associated with minimal hypersurfaces, $X_{K}$ $=0$, of thermodynamic systems
of Pfaff topological dimension 4. \ 

\begin{center}
The 26 kb color presentation of Figure 1 can be downloaded from

http://www22.pair.com/csdc/pdf/univgibb.jpg

\textbf{Fig 1. \ Universal Topological Gibbs function.}
\end{center}

The similarity coordinate $T_{K}$ plays the role of the Gibbs free energy, in
terms of the Pressure $(\symbol{126}Z_{A})$ and the Temperature $(\symbol{126}%
Y_{G})$. \ The Spinodal line as a limit of phase stability, and the critical
point are ideas that come from the study of a van der Waals gas, but herein it
is apparent that these concepts are universal topological concepts that remain
invariant with respect to deformations. \ The universal formulas for such
constraints are presented in the next section. \ The result is that all
thermodynamic systems of Pfaff topological dimension 4 are deformably
equivalent to a van der Waals gas.

It is important to recognize that the development of a universal
nonequilibrium van der Waals gas has not utilized the concepts of metric,
connection, statistics, relativity, gauge symmetries, or quantum mechanics.

\subsubsection{The Edge of Regression and Self Intersections}

The envelope is smooth as long as $\partial^{2}\Theta/\partial\Psi^{2}%
=\Theta_{\Psi\Psi}\neq0,$ and $d\Theta\symbol{94}d\Theta_{\Psi}\neq0,$ subject
to the further constraint that the family parameter is a constant: $d\Psi=0.$
\ If $d\Theta\symbol{94}d\Theta_{\Psi}\neq0$, but $\Theta_{\Psi\Psi}=0,$ then
the envelope has a self intersection singularity. \ If $d\Theta\symbol{94}%
d\Theta_{\Psi}=0$, but $\Theta_{\Psi\Psi}\neq0,$ there is no self
intersection, and no envelope. \ 

If the envelope exists, further singularities are determined by the higher
order partial derivatives of the Universal Phase function with respect to
$\Psi$. \
\begin{equation}
\partial^{2}\Theta/\partial\Psi^{2}=\Theta_{\Psi\Psi}=12\Psi^{2}-6X_{M}%
\Psi+2Y_{G}.
\end{equation}%
\begin{equation}
\partial^{3}\Theta/\partial\Psi^{3}=\Theta_{\Psi\Psi\Psi}=24\Psi-6X_{M}%
\end{equation}
When $\partial^{3}\Theta/\partial\Psi^{3}=\Theta_{\Psi\Psi\Psi}\neq0,$ and
$d\Theta\symbol{94}d\Theta_{\Psi}\symbol{94}d\Theta_{\Psi\Psi}\neq0$, the
envelope terminates in a edge of regression. \ The edge of regression is
determined by the simultaneous solution of \ $\Theta=0,\Theta_{\Psi}=0$ and
$\Theta_{\Psi\Psi}=0.$ \ For the minimal surface representation of the Gibbs
surface for a van der Waals gas, the edge of regression defines the Spinodal
line of ultimate phase stability. \ The edge of regression is evident in the
Swallowtail figure\ (above) describing the Gibbs function for a van der Waals
gas. \ 

If $\Theta_{\Psi\Psi\Psi}=0,$ then for $X_{M}=0,$ it follows that
$Y_{G}=0,\ Z_{A}=0,\ T_{K}=0,$ which defines the critical point of the \ Gibbs
function for the van der Waals gas. \ In other words, the critical point is
the zero of the 4-dimensional space of similarity coordinates.

If $\Theta_{\Psi\Psi}=0,$ then for $X_{M}=0$ the envelope has a self
intersection. \ It follows from $\Theta_{\Psi\Psi}=0,$ that $\Psi^{2}%
=-Y_{G}/6,$ which when substituted into%

\begin{equation}
\Theta_{\Psi}=4\Psi^{3}+2Y_{G}\Psi-Z_{A}\Rightarrow0,
\end{equation}
yields the%

\begin{equation}
\text{Universal Gibbs Edge of Regression:\ }Z_{A}^{2}+Y_{G}^{3}(8/27)=0,
\end{equation}
which defines the Spinodal line, of the minimal surface representation for a
universal nonequilibrium van der Waals gas, in terms of
"similarity"\ coordinates. \ 

\ Within the swallow tail region the "Gibbs" surface has 3 real roots and
outside the swallow tail region there is a unique real root. \ The edge of
regression furnished by the Cardano function defines the transition between
real and imaginary root structures. \ The details of the universal
nonequilibrium van der Waals gas in terms of envelopes and edges of regression
with complex molal densities or order parameters will be presented elsewhere.
\ These systems are not equilibrium systems for the Pfaff dimension is not 2.
\ \ Of obvious importance is the idea that the a zero value for both $Z_{G}$
and $T_{K}$ are required to reduce the Pfaff dimension to 2, which is the
necessary condition for an isolated or equilibrium system.

\subsection{Ginsburg Landau Currents}

The Universal Phase function can be solved for the determinant of the Jacobian
matrix, which is equal to the similarity invariant $T_{K},$%

\begin{equation}
T_{K}=-\{\Psi^{4}-X_{M}\Psi^{3}+Y_{G}\Psi^{2}-Z_{A}\Psi\}.
\end{equation}
All determinants are, in effect, N - forms on the domain of independent
variables. \ All N-forms can be related to the exterior differential of some
N-1 form or current, $J.$ \ Hence%

\begin{equation}
dJ=K\Omega_{4}=div\mathbf{J}+\partial\rho/\partial t=-(\Psi^{4}-X_{M}\Psi
^{3}+Y_{G}\Psi^{2}-Z_{A}\Psi)\Omega_{4}.
\end{equation}
For currents of the form%

\begin{align}
\mathbf{J}  &  =grad\ \Psi,\\
\rho &  =\Psi,
\end{align}
the Universal Phase function generates the universal Ginsburg Landau equations%

\begin{equation}
\nabla^{2}\Psi+\partial\Psi/\partial t=-(\Psi^{4}-X_{M}\Psi^{3}+Y_{G}\Psi
^{2}-Z_{A}\Psi),.
\end{equation}
and establishes contact with the "bottom up" methods.

\subsection{Singularities as defects of Pfaff dimension 3}

The family of hypersurfaces can be topologically constrained such that the
topological dimension is reduced, and/or constraints can be imposed upon
functions of the similarity variables forcing them to vanish. \ Such regions
\ in the 4 dimensional topological domain indicate topological defects or
thermodynamic changes of phase. \ It is remarkable that for a given 1-form of
Action there are an infinite number rescaling functions, $\lambda,$ such that
the Jacobian matrix $\left[  \mathbb{J}_{jk}^{scaled}\right]  =[\partial
(A/\lambda)_{j}/\partial x^{k}]$ is singular (has a zero determinant). \ \ For
if the coefficients of any 1-form of Action are rescaled by a divisor
generated by the Holder norm,
\begin{equation}
\text{Holder Norm: \ }\lambda=\{a(A_{1})^{p}+b(A_{2})^{p}+c(A_{3})^{p}%
+e(A_{4})^{p}\}^{m/p},\label{holder}%
\end{equation}
then the rescaled Jacobian matrix%

\begin{equation}
\left[  \mathbb{J}_{jk}^{scaled}\right]  =[\partial(A/\lambda)_{j}/\partial
x^{k}]
\end{equation}
will have a zero determinant, for any index p, any set of isotropy or
signature constants, a, b, c, e, if the homogeneity index is equal to unity:
$m=1$. \ This homogeneous constraint implies that the similarity invariants
become projective invariants, not just equi-affine invariants. \ Such species
of topological defects can have the image of a 3-dimensional implicit
characteristic hypersurface in space-time:%

\begin{equation}
\text{Singular hypersurface in 4D: \ }\det[\partial(A/\lambda)_{j}/\partial
x^{k}]\Rightarrow0
\end{equation}
The singular fourth order Cayley-Hamilton polynomial of $\left[
\mathbb{J}_{jk}\right]  $ then will have a cubic polynomial factor with one
zero eigenvalue. \ 

For example, consider the simple case where the determinant of the Jacobian
vanishes: $T_{K}\Rightarrow0.$ \ Then the Phase function becomes%
\begin{align}
\text{Universal Equation of State}  &  \text{: \ }\Theta(\{X_{M},Y_{G}%
,Z_{A},T_{K}=0\};\ \Psi)\label{pvt}\\
&  =\Psi(\Psi^{3}-X_{M}\Psi^{2}+Y_{G}\Psi-Z_{A})\Rightarrow0.
\end{align}
The space has been topologically reduced to 3 dimensions (one eigen value is
zero), and the zero set of the resulting singular Universal Phase function
becomes a universal cubic equation that is homeomorphic to the cubic equation
of state for a van der Waals gas.

When the rescaling factor $\lambda$ is chosen such that $p=2,a=b=c=1,m=1$,
then the Jacobian matrix, $\left[  \mathbb{J}_{jk}\right]  ,$ is equivalent to
the "Shape" matrix for an implicit hypersurface in the theory of differential
geometry. \ Recall that the homogeneous similarity invariants can be put into
correspondence with the linear Mean curvature, $X_{M}\Rightarrow C_{M}$, the
quadratic Gauss curvature, $Y_{G}\Rightarrow C_{G}$, and the cubic Adjoint
curvature, $Z_{A}\Rightarrow C_{A},$ of the hypersurface. \ The characteristic
cubic polynomial can be put into correspondence with a nonlinear extension of
an ideal gas \textit{not necessarily} in an equilibrium state. \ 

\subsection{The Universal van der Waals gas}

More that 100 years ago van der Waals introduced into the science of
thermodynamics the equation of state now called the van der Waals gas:%

\begin{equation}
P=\rho RT/(1-b\rho)+a\rho^{2}%
\end{equation}
The van der Waals equation may be considered as a cubic constraint on the
space of variables $\{n;P,V,T\}$ where $\rho=n/V$ is defined as the molar density.%

\begin{equation}
\rho^{3}-(1/b)\rho^{2}+\{-(RT+bP)/ab\}\rho+P/ab=0.
\end{equation}
This cubic equation is to be compared with the characteristic polynomial
written in terms of the similarity invariants, $M,\ G,\ $and $A.$ \ Note that
the roots of the characteristic polynomial are not necessarily real. \ This
observation leads to a well defined procedure for treating nonequilibrium
thermodynamics systems as complex deviations from the real, or equilibrium,
systems. \ The reality condition is determined by the Cardano function that
describes an edge of regression discontinuity. \ 

For a transformation such that
\begin{equation}
(8T+P)/3=Y_{G}/(M/3)^{2},
\end{equation}%
\begin{align}
P  &  =Z_{K}/(M/3)^{3},\\
\lambda &  =-\rho/(M/3),
\end{align}
the characteristic polynomial becomes an equation in terms of dimensionless
parameters,%
\begin{equation}
U(\lambda,T,P)=(\lambda)^{3}-3(\lambda)^{2}+[(8T+P)/3](\lambda)-P=0.
\end{equation}
The last format given above is to be recognized as the Equation of State of a
van der Waals Gas (compare to \ref{pvt}), in terms of dimensionless Pressure,
Temperature relative to their values at the critical point. \ \ 

\section{The Falaco Cosmological Soliton}

Although of importance to the cosmological concept of a universe expressible
as a low density (nonequilibrium)\ van der Waals gas near its critical point,
the factorization of the Jacobian characteristic polynomial into a cubic is
not the only cosmological possibility. \ Of particular interest is the
factorization that leads to a Hopf bifurcation. \ In this case the
characteristic determinant vanishes, the Adjoint cubic curvature vanishes, the
mean curvature vanishes (indicating a minimal surface), but the Gauss
curvature is positive, and the two remaining eigenvalues of the characteristic
polynomial are pure imaginary conjugates. \ Such results indicate rotations or
oscillations (as in the chemical Brusselator reactions) and the possibility of
spiral concentration or density waves on such minimal surfaces. \ Such
structures at a cosmological level would appear to explain the origin of
spiral arm galaxies. \ The Hopf type minimal surfaces of positive Gauss
curvature do not represent thermodynamic equilibrium systems, for their
curvatures, although two in number, are pure imaginary. \ The molal density
distributions (or order parameters) are complex.

Evidence\ of such topological defects (at the macroscopic level) can be
demonstrated by the creation of Falaco Solitons in a swimming pool
\cite{rmkfalaco} \cite{vol2}. (See Figure 2). \ 

\begin{center}
The 54 kb color photo of Figure 2 can be downloaded from

http://www22.pair.com/csdc/pdf/falcolor.jpg

\textbf{Fig 2. \ Falaco Solitons}

\textbf{Cosmic strings in a swimming pool}
\end{center}

These experiments demonstrate that such topological defects are available at
all scales. \ The Falaco Solitons consist of spiral "vortex defect" structures
(analogous to CGL theory)\ on a two dimensional minimal surface, one at each
end of a 1-dimensional "vortex line" or thread (analogous to GPG theory).
\ Remarkably the topological defect surface structure is locally unstable, as
the surface is of negative Gauss curvature. \ Yet the pair of locally unstable
2-D surfaces is \textit{globally} stabilized by the 1-D line defect attached
to the "vertex" points of the minimal surfaces. \ It is remarkable to me that
the Falaco Solitons are obvious repeatable experimental examples of "strings
connected to branes", yet no string theorist that I have challenged to show
how his string "theory" describes the emergence of Falaco Solitons has
responded with a solution. \ My view is that I hold the fanciful claims\ of
string theory suspect, until those theorists can demonstrate a solution that
describes the experimental Falaco Solitons. \ 

\begin{center}
The 26 kb color presentation of Figure 3 can be downloaded from

http://www22.pair.com/csdc/pdf/tornqr3.jpg

\textbf{Fig 3. \ Falaco Solitons and Landau Ginsburg theory.}
\end{center}

For some specific physical systems it can be demonstrated that period
(circulation) integrals of the 1-form of Action potentials, $A,$ lead to the
concept of "vortex defect lines". \ The idea is extendable to "twisted vortex
defect lines" in three dimensions. \ The "twisted vortex defects" become the
spiral vortices of a Complex Ginsburg Landau (CGL) theory , while the
"untwisted vortex lines" become the defects of Ginzburg-Pitaevskii-Gross (GPG)
theory \cite{Tornkvist}. \ In my opinion, it is unfortunate that the word
"vortex" has been used so glibly in such descriptive phrases. \ To a fluid
dynamicist, the concept of a vortex implies the existence of vorticity (curl
of the velocity field). \ Circulation is a fluid property independent from the
existence of vorticity. \ I\ suggest that the descriptions "Vortex defect
lines" should be made more precise in terms of the phrase "Circulation defect lines".

\ In the macroscopic domain, the fluid experiments visually indicate the
emergence of "almost flat" spiral arm structures during the formative stages
of the Falaco solitons. \ In the cosmological domain, it is suggested that
these universal topological defects represent the ubiquitous "almost flat"
spiral arm galaxies. \ Based on the experimental creation of Falaco Solitons
in a swimming pool, it has been conjectured that M31 and the Milky Way
galaxies could be connected by a topological defect thread \cite{rmkfalaco}.
\ Only recently has photographic evidence appeared suggesting that galaxies
may be connected by "strings" (Figure 4).

\begin{center}
The 26 kb Hubble photo\ of Figure 4 can be downloaded from

http://www22.pair.com/csdc/pdf/spiralstring.jpg

\textbf{Fig 4. \ Interacting Spiral Galaxies}
\end{center}

At the other extreme, using drops of dye, the rotational minimal
surfaces\footnote{In euclean 3-space such minimal surfaces have a negative
Gauss curvature, but in Minkowski 3 space they have positive Gauss curvature.}
which form the two endcaps of the Falaco soliton, like quarks, apparently are
confined by a "string". If the "string" (whose "tension" induces global
stability of the unstable endcaps) is severed, the endcaps (like unconfined
quarks in the elementary particle domain)\ disappear (in a non-diffusive
manner). \ In the microscopic electromagnetic domain, the Falaco soliton
structure offers an alternate, topological, pairing mechanism on a Fermi
surface, that could serve as an alternate to the Cooper pairing in superconductors.

\section{The Adjoint Current and Topological Spin}

From the singular Jacobian matrix, $\left[  \mathbb{J}_{jk}^{scaled}\right]
=[\partial(A/\lambda)_{j}/\partial x^{k}],$ it is always possible to construct
the Adjoint matrix as the matrix of cofactors transposed: \
\begin{equation}
\text{Adjoint Matrix : }\left[  \widehat{\mathbb{J}}^{kj}\right]
=adjoint\left[  \mathbb{J}_{jk}^{scaled}\right]
\end{equation}
When this matrix is multiplied times the rescaled covector components, the
result is the production of an adjoint current,%

\begin{equation}
\text{Adjoint current : }\left\vert \widehat{\mathbf{J}}^{k}\right\rangle
=\left[  \widehat{\mathbb{J}}^{kj}\right]  \circ\left\vert \mathbf{A}%
_{j}/\lambda\right\rangle
\end{equation}
It is remarkable that the construction is such that the Adjoint current
3-form, if not zero, has zero divergence globally \cite{rmkpoincare}:%

\begin{align}
\widehat{J}  &  =i(\widehat{\mathbf{J}}^{k})\Omega_{4}\\
d\widehat{J}  &  =0.
\end{align}
From the realization that the Adjoint matrix may admit a nonzero globally
conserved 3-form density, or current, $\widehat{J},$ it follows abstractly
that there exists a 2-form density of "excitations", $\widehat{G},$ such that%

\begin{equation}
\text{Adjoint current : }\widehat{J}\Leftarrow d\widehat{G}.
\end{equation}
$\widehat{G}$ is not uniquely defined in terms of the adjoint current, for
$\widehat{G}$ could have closed components (gauge additions $\widehat{G}_{cl},
$ such that $d\widehat{G}_{cl}=0$), which do not contribute to the current,
$\widehat{J}.$

From the topological theory of electromagnetism \cite{rmktop} \cite{rmk4eyes1}%
\ there exists a fundamental 3-form, $A\symbol{94}\widehat{G},$ defined as the
"topological Spin" 3-form,%

\begin{equation}
\text{Topological Spin 3-form \ : \ }A\symbol{94}\widehat{G}.
\end{equation}
The exterior differential of this 3-form produces a 4-form, with a coefficient
energy density function that is composed of two parts:%

\begin{equation}
d(A\symbol{94}\widehat{G})=F\symbol{94}\widehat{G}-A\symbol{94}\widehat{J}.
\end{equation}
The first term is twice the difference between the "magnetic" and the
"electric" energy density, and is a factor of 2 times the Lagrangian usually
chosen for the electromagnetic field in classic field theory:
\begin{equation}
\text{Lagrangian Field energy density : }F\symbol{94}\widehat{G}%
=\mathbf{B\circ\,H-D\circ E}%
\end{equation}
The second term is defined as the "interaction energy density"%

\begin{equation}
\text{Interaction energy density : }A\symbol{94}\widehat{J}=\mathbf{A\circ
}\widehat{\mathbf{J}}-\rho\phi.
\end{equation}
For the special (Gauss)\ choice of integrating denominator, $\lambda$ with
$(p=2,a=b=c=1,m=1)$ it can be demonstrated that the Jacobian similarity
invariants are equal to the classic curvatures:%
\begin{equation}
\{X_{M},Y_{G},Z_{A},T_{K}\}\Rightarrow\{C_{M(mean\_linear)}%
,C_{G(gauss\_quadratic)},C_{A(adjoint\_cubic)},0\}.
\end{equation}
\ It can be demonstrated further that the interaction density is exactly equal
to the Adjoint curvature energy density:
\begin{equation}
\text{Interaction energy }A\symbol{94}\widehat{J}=C_{A}\ \Omega_{4}\text{
\ \ (The Adjoint Cubic Curvature).}%
\end{equation}
The conclusion reached is that a nonzero interaction energy density implies
the thermodynamic system is not in an equilibrium state. \ 

\ However, it is always possible to construct the 3-form, $\widehat{S}:$%

\begin{equation}
\text{Topological Spin 3-form : }\widehat{S}=A\symbol{94}\widehat{G}%
\end{equation}
The exterior differential of this 3-form leads to a cohomological structural
equation similar the first law of thermodynamics, but useful for
nonequilibrium systems. \ This result, now recognized as a statement
applicable to nonequilibrium thermodynamic processes, was defined as the
"Intrinsic Transport Theorem" in 1969 \cite{rmkintrinsic} :%

\begin{align}
\text{Intrinsic Transport Theorem (Spin)} &  :\ \ \ d\widehat{S}%
=F\symbol{94}\widehat{G}-A\symbol{94}\widehat{J},\\
\text{First Law of Thermodynamics (Energy)} &  :dU=Q-W
\end{align}
If one considers a collapsing system, then the geometric curvatures increase
with smaller scales. \ If Gauss quadratic curvature, $C_{G},$ is to be related
to gravitational collapse of matter, then at some level of smaller scales a
term cubic in curvatures, $C_{H},$ would dominate. \ It is conjectured that
the cubic curvature produced by the interaction energy effect described above
could inhibit the collapse to a black hole. \ \ Cosmologists and relativists
apparently have ignored such cubic curvature effects.

\section{Topological Fluctuations}

\ Topological fluctuations are admitted when the evolutionary vector direction
fields are not singly parametrized:%

\begin{align}
\text{Fluctuations in position (pressure) }\text{: } &  d\mathbf{x}%
-\mathbf{v}dt=\Delta\mathbf{x}\neq0\\
\text{Fluctuations in velocity (temperature) }\text{: } &  d\mathbf{v}%
-\mathbf{a}dt=\Delta\mathbf{v}\neq0\\
\text{Fluctuations in momenta (viscosity) }\text{: } &  d\mathbf{p}%
-\mathbf{f}dt=\Delta\mathbf{p}\neq0.
\end{align}
These failures of kinematic perfection undo the topological refinements
imposed by a "kinematic particle" point of view, and place emphasis on the
continuum methods inherent in fluids and plasmas. \ For example, consider the
Cartan-Hilbert 1-form of Action on a space of 3n+1 independent
variables\footnote{The domain of independent variables is not restricted to
dimension 4 in this section.} (the $p_{\mu}$ are presumed to be independent
Lagrange multipliers):%

\begin{equation}
A=L(\mathbf{x},\mathbf{v},t)dt+p_{\mu}(dx^{\mu}-v^{\mu}dt)=L(x,v,t)dt+p_{\mu
}\Delta x^{\mu})
\end{equation}
The Top Pfaffian in the Pfaff sequence is%

\begin{equation}
(dA)^{n+1}=(n+1)!\{\Sigma_{\mu=1}^{n}(\partial L/\partial v^{\mu}-p_{\mu
})\bullet d\mathbf{v}^{\mu}\}\symbol{94}dp_{1}\symbol{94}...dp_{n}%
\symbol{94}dq^{1}\symbol{94}..dq^{n}\symbol{94}dt,
\end{equation}
and yields a Pfaff dimension of 2n+2 for the 1-form of Action, defined on the
geometric space of 3n+1 variables $\{x^{\mu},p_{\mu},v^{\mu},t\}$. \ This even
dimensional space defines a symplectic manifold. \ 

For the maximal non-canonical symplectic physical system of Pfaff dimension
2n+2, consider evolutionary processes to be representable by vector fields of
the form $\gamma V_{3n+1}=\gamma\{\mathbf{v,a,f},1\},$ relative to the
independent variables $\{\mathbf{x,v,p},t\}.$ Define the \textquotedblright
virtual work\textquotedblright\ 1-form, $W$, as $W=i(\mathbf{W})dA$, a 1-form
which must vanish for the extremal case, and be nonzero, but closed, for the
symplectic case. For any n, it may be shown by direct computation that the
virtual work 1-form consists of two distinct terms, each involving a different
fluctuation:%
\begin{equation}
W=\{\mathbf{p}-\partial L/\partial\mathbf{v}\}\bullet\Delta\mathbf{v}%
+\{\mathbf{f}-\partial L/\partial\mathbf{x}\}\bullet\Delta\mathbf{x}%
\end{equation}
When the fluctuations in velocity are zero (temperature) and the fluctuations
in position are zero (pressure), then the work 1-form will vanish, and the
process and physical system admits a Hamiltonian representation. \ On the
other hand if the fluctuations in velocity are not zero and the fluctuations
in position are not zero, then the Work 1-form vanishes only if the momenta
(the Lagrange multipliers, $\mathbf{p,}$ are canonically defined
($\{\mathbf{p}-\partial L/\partial\mathbf{v}\}\Rightarrow0$) and the Newtonian
force is a gradient, \ $\{\mathbf{f}-\partial L/\partial\mathbf{x}%
\}\Rightarrow0.$ \ These topological constraints are ubiquitously assumed in
classical mechanics.

When $\Delta x^{k}\Rightarrow0,$ such that all topological fluctuations
vanish, then the Pfaff dimension of the physical system defined in terms of
the Cartan-Hilbert 1-form of Action, $A,$ is 2 (the equilibrium requirement).

\section{Examples of thermodynamic 1-forms}

In order to demonstrate content to the thermodynamic topological theory, two
algebraically simple examples are presented below. \ The first corresponds to
a Jacobian characteristic equation that has a cubic polynomial factor, and
hence can be identified with a van der Waals gas. \ The second example
exhibits the features associated with a Hopf bifurcation, where the
characteristic equation has a quadratic factor with two pure imaginary roots,
and two null roots. \ Another example, given in \cite{vol2}, \cite{rmkarw},
demonstrates how a bowling ball, given initial angular momentum and energy,
skids and/or slips changing its angular momentum and kinetic energy
irreversibly via friction effects, until the dynamics is such that the ball
rolls with out slipping. \ Once that "excited" state is reached, and
topological fluctuations are ignored, the motion continues without
dissipation. \ The system is in an excited state far from equilibrium. \ 

\subsection{Example 1: van der Waals properties from rotation and contraction}

\vspace{1pt}In this example,the Action 1-form is presumed to be of the form
\begin{equation}
A_{0}=a(ydx-xdy)+b(tdz+zdt).
\end{equation}
The 1-form of Potentials depends on the coefficients $a$ and $b.$ \ The
results of the topological theory are (for $r^{2}=x^{2}+y^{2}+z^{2}+t^{2})$:
\begin{align}
\text{ Mean curvature}  &  \text{:}C_{M}=-2btz/(r^{2})^{3/2}\\
\text{Gauss curvature}  &  \text{:}C_{G}=-\{b^{2}(x^{2}+y^{2})-a^{2}%
(z^{2}+t^{2})\}/(r^{2})^{2}\\
\text{ Adjoint curvature}  &  \text{:}C_{A}=A\symbol{94}J_{s}\;=-2a^{2}%
btz/(r^{2})^{5/2}\\
Top\_Torsion  &  =2ab\ \cdot\lbrack0,0,z,-t]/(r^{2})\\
\text{ Adjoint Current \ \ }  &  \text{:}J_{s}=(a^{2}b^{2}\cdot\lbrack
x,y,z,t])\ /(r^{2})^{2}\\
\text{Pfaff Dimension 4}  &  \text{:}dA\symbol{94}dA=2ba(t^{2}-z^{2}%
)/(r^{2})^{2}\ \Omega_{4}%
\end{align}
The Jacobian matrix has 1 zero eigen value and three nonzero eigenvalues.
\ Hence, the cubic polynomial will yield an interpretation as a van der Waals
gas. \ The Adjoint current represents a contraction in space-time, while the
flow associated with the 1-form has a rotational component about the z axis.

\subsection{\vspace{1pt}\vspace{1pt}Example 2: A Hopf 1-form \ }

\vspace{1pt}In this example,the Hopf 1-form is presumed to be of the form
\begin{equation}
A_{0}=a(ydx-xdy)+b(tdz-zdt).
\end{equation}
The 1-form of Potentials depends on the coefficients $a$ and $b.$ \ There are
two cases corresponding to left and right handed \textquotedblright
polarizations\textquotedblright: \ $a=b$ or $a=-b$. \ The results of the
topological theory are (for $r^{2}=x^{2}+y^{2}+z^{2}+t^{2})$:
\begin{align}
\text{ Mean curvature}  &  \text{:}C_{M}=0,\\
\text{Gauss curvature}  &  \text{:}C_{G}=\{b^{2}(x^{2}+y^{2})+a^{2}%
(z^{2}+t^{2})\}/(r^{2})^{2}\\
\text{ Adjoint Cubic curvature}  &  \text{:}C_{A}=A\symbol{94}J_{s}\;=0\\
Top\_Torsion  &  =2ab\ \cdot\lbrack x,y,z,t]/(r^{2})\\
\text{ Adjoint Current \ \ }  &  \text{:}J_{s}=(ab/2)\ \cdot Top\_Torsion\\
\text{Pfaff Dimension 4}  &  \text{:}dA\symbol{94}dA=4ab/(r^{2})\ \Omega_{4}%
\end{align}

What is remarkable for this Action 1-form is that both the mean curvature and
the Adjoint curvature of the implicit hypersurface in 4D vanish, for any
choice of a or b. \ The Gauss curvature is nonzero, positive real and is equal
to the inverse square of the radius of a 4D euclidean sphere, when
$a^{2}=b^{2}$. \ The Adjoint cubic interaction energy density is zero. \ The
two nonzero curvatures are pure imaginary conjugates equal to
\begin{equation}
\ \rho=\pm\sqrt{-b^{2}(x^{2}+y^{2})-a^{2}(z^{2}+t^{2})}/(r^{2}).\
\end{equation}

\ The Hopf surface is a 2D imaginary \textit{minimal} two dimensional hyper
surface in 4D and has two nonzero imaginary curvatures! \ Strangely enough the
charge-current density is not zero, but it is proportional to the Topological
Torsion vector that generates the 3 form $A\symbol{94}F.$ \ The topological
Parity 4 form is not zero, and depends on the sign of the coefficients a and
b. \ In other words the 'handedness' of the different 1-forms determines the
orientation of the normal field with respect to the implicit surface. \ It is
known that a process described by a vector proportional to the topological
torsion vector in a domain where the topological parity is nonzero
$4ba/(x^{2}+y^{2}+z^{2}+t^{2})$ $\neq0$ is thermodynamically irreversible. \ 

\subsection{Example 3 \ A repeatable experiment that demonstrates emergence}

As an example that can be experimentally replicated regard the photo below
(Figure 5). \ The fascinating thing to me\footnote{In 1957 as I stood in the
Yucca Flats valley of Nevada} was how, in the midst of all the turbulent
irreversible dissipation (Pfaff topological dimension 4) associated with a
nuclear explosion, there would emerge a topologically coherent, nonequilibrium
macroscopic state that was radiating (Pfaff topological dimension 3) in the
form of a toroidal topological defect. \ A surprising observation was that
this excited nonequilibrium state had relative long lifetime.

\begin{center}
The 36 kb Color Photo of Fig 5 can be downloaded from

http://www22.pair.com/csdc/pdf/priscila.jpg

\textbf{Figure 5} \textbf{Ionized toroidal topological defect}

\textbf{Pfaff topological dimension 3}
\end{center}

\section{Conclusions Part I}

Based upon the single assumption that the universe is a nonequilibrium
thermodynamic system of Pfaff topological dimension 4 leads to a cosmology
where the universe, at present, can be approximated in terms of the
nonequilibrium states of a very dilute van der Waals gas near its critical
point. \ The stars and the galaxies are the topological defects and coherent
(but not equilibrium) self-organizing structures of Pfaff topological
dimension 3 formed by irreversible topological evolution in this
nonequilibrium system of Pfaff topological dimension 4.

The turbulent nonequilibrium thermodynamic cosmology of a real gas near its
critical point yields an explanation for:

\begin{enumerate}
\item The granularity of the night sky as exhibited by stars and galaxies.

\item The Newtonian law of gravitational attraction proportional to 1/r$^{2}.
$

\item The expansion of the turbulent dissipative universe.

\item The emergence of nonequilibrium (radiating) Pfaff dimension 3,
topological defect structures such as stars and galaxies.

\item A possible understanding of non radiating systems (dark energy, dark
matter) in terms of ordinary thermodynamic defect systems of Pfaff topological
dimension less than 3. \ Such thermodynamic systems do not exchange matter or
radiative energy with the turbulent dissipative environment of the physical
cosmological vacuum.
\end{enumerate}

\begin{quote}
The color photos described above in Part I have been presented in a somewhat
awkward manner as there is an arXiv limit on file size. \ A pdf file (1.46 Mb)
that includes the color photos in place can be downloaded from
\end{quote}

\begin{center}
http://www22.pair.com/csdc/pdf/coscolor.pdf
\end{center}

\part{Macroscopic Topological Quantization}

Part II examines how continuous topological evolution can be used to describe
the thermodynamic emergence of topological defect singular structures without
regard to geometric scales. \ Moreover, these deformable, but topologically
coherent, signular structures can exhibit macroscopic, topologically
quantized, (rational) properties which can be used to describe the features of
quantum cosmology. \ The bottom line is the idea that Quantum Cosmology should
be treated as a topological, not a metrical, concept. \ The work is motivated
by the conjecture that the cosmology of the observable universe can be
described as a dilute, but turbulent, thermodynamic state of Pfaff topological
dimension 4. \ Irreversible thermodynamic processes cause the emergence of
various regional defect domains (such as condensates) of coherent, but
deformable, topological features, of Pfaff topological dimension 3, or less.
\ Tangential discontinuities such as wakes in fluids, and propagating
electromagnetic signals are examples of such emergent singular topological
defects. \ 

In addition, certain homogeneous defect structures, which can occur over
microscopic or cosmological domains, admit features of quantization in terms
of deRham period integrals, which are known to have rational values. \ The
homogeneous structures introduce singularities of many different forms into
the topological background. \ The simplest of these structures are related to
fixed points of rotation and expansion. \ 

\section{Emergence}

During the last 5 years, or so, the old concept of "emergent physics" has
developed into a buzzword that attracts attention in the scientific community.
\ From a topological perspective, the word \textit{emergence}, describing a
process that causes something "new" to be observed could have two
interpretations.\ \ Both involve topological change. \ 

\begin{enumerate}
\item The first suggestion is associated with the idea of creation; the
emergence of something as a "new" entity (or final state), with topological
properties\footnote{Note that herein the emphasis is on topological
properties, not geometric properties. \ Metric features, more or less, are
ignored.} that are different from preceding entity (or initial state).
\ Separation of a checkerboard into its many black parts and its many red
parts is an elementary example of a "cutting" process. \ Indeed, the
topological property of connectivity has changed during the process, but the
(cutting)\ process cannot be represented by a topologically continuous
mapping. \ The disconnected pieces are the "new" entity that emerges after the
cutting process takes place. \ 

\item The second suggestion is based on the observation that it is possible to
start with a strip of ribbon, which is simply connected and orientable, and
cause it to emerge (evolve) into a "new" entity which is not orientable and
not simply connected. \ This "twisting" and "pasting" process does indeed
describe topological change, but the (twisting and pasting)\ process can be
represented by a topologically continuous mapping. \ 
\end{enumerate}

Other examples include the emergence of a turbulent state, from an equilibrium
state of rest; \ but such a process can not be described in terms of a
topologically continuous process. \ \ On the other hand the decay of a
turbulent state, to state of rest, does involve topological change that can be
mapped by a continuous process. \ In general, it is known that a physical
system encoded by a connected topology can not continuously be mapped (evolve)
into a disconnected topology. \ On the other, a disconnected topology can
continuously be mapped (evolve) into connected topology. \ 

\ For systems encoded in terms of a Cartan 1-form, the induced Cartan topology
is such that the domains of Pfaff topological dimension 1 or 2 form connected
topologies (which are representative of thermodynamic equilibrium, or
thermodynamic isolated states), and the domains of Pfaff topological dimension
3 or 4 form disconnected topologies (which are representative of thermodynamic
states far-from-equilibrium or thermodynamic turbulent states). \ These ideas
do not depend upon metric scales. \ 

The focus herein is on continuous topological evolution, for which the use of
Cartan's exterior calculus will lead to progress in scientific understanding.
\ Discontinuous\footnote{Topological continuity requires that the limit points
of the topology of the initial state map into the closure of the
(different)\ topology of the final state. \ Topology can change continuously.}
topological evolution is ignored herein. \ \ The turbulent state on a
pregeometric variety (no metric) \ of 4 variables is defined to be of Pfaff
topological dimension 4. \ The initial turbulent state can decay (or evolve)
continuously into macroscopic topological coherent structures of lesser Pfaff
topological dimension. \ These emergent structures can be considered to be
topological defects in the domain of Pfaff topological dimension 4. \ Those
emergent states (domains) that are of Pfaff dimension 3, created from
dissipative irreversible processes in the turbulent environment (domain) of
Pfaff topological dimension 4, are topological defect structures that can have
remarkably long lifetimes. \ They are not thermodynamic equilibrium states,
for they have a Pfaff topological dimension 3.

Limit cycles, envelopes, excited atomic states, Solitons, wakes, galaxies,
envelopes, stars are all examples of such topologically coherent but
deformable structures of Pfaff topological dimension 3. \ Perhaps even more
remarkable is the idea that these coherent topological defect structures, if
homogeneous, can be used to generate topological quantization effects that are
not dependent upon scales. \ Such macroscopic "quantum" states can occur at
the size of galaxies as well as at the size of Bose condensates. \ 

An interesting experiment relating to the concept of irreducible Pfaff
dimension 3, non-equilibirum thermodynamics, topological quantum states, and
the fact that a twisting and pasting continuous process can store energy in
physical systems by means of curvature and torsion can be conducted by using a
length of thick wall elastic vacuum hose. \ Bend the hose into a circle and
join the ends together without twisting. \ The curvature deformation of
compression of the inside fibers,and extension of the outside fibers required
work to be done. \ The stored energy of deformation can be retrieved. \ If the
hose is placed on a table top it lies flat; the deformed fibers reside in a
plane. \ Now before joining the ends together give the hose a pi twist.
\ There is now obvious deformation energy associated with the curvature, but
there also is an additional energy associated with the twist or torsion
deformation. \ Place the hose with both curvature and torsion on the table
top. \ It does not lie flat. \ It is irreducibly 3 dimensional as it has torsion.

An open question is: Does the physical vacuum have this torsional energy;
\ can it be retrieved? \ If the physical vacuum is\ described by a matrix of
Basis vector functions, then it appears that the Affine torsion of the
associated Cartan connection leads to PDE's of the format equivalent to both
the Maxwell-Ampere equations and the Maxwell-Faraday equations. \ The
conclusion is reached that the source of charge is Affine Torsion of the
Cartan Connection Matrix constructed from the Basis Frame of Functions that
define the Physical Vacuum \cite{rmkpv}. \ Of particular interest is the
theoretical conjecture that Affine torsion of the Cartan Connection for the
physical vacuum is the source of charge.\bigskip

\subsection{Historical}

More than 25 years ago (1977), the present author published an article
entitled,\ "Periods on Manifolds, Quantization and Gauge"\ \cite{rmkperiods}.
\ At that time, it had become apparent that at least some of the quantum
mechanical features of measurables with rational ratios (the quantum numbers)
could be interpreted in terms of topological period integrals (which have
ratios of values that are rational). \ The method was championed then and now
by E. J. Post \cite{PostQuRe}, who, using the methods of topological
quantization, predicted in 1981 the fractional quantum Hall effect
\cite{PostQHall}. \ Further motivation for the original publication was based
on the recognition that the evolution of the Sommerfeld closed integrals might
be used to explain the details of that Copenhagen mystery, whereby the quantum
jump, or radiative transition from one quantum state to another quantum state,
had been described (paraphrasing Bohr) as a "miracle". \ It was recognized
that a quantum transition, which was described by integer changes of the
Sommerfeld Period integrals, implied a topological change had taken place.

A few years earlier it had been realized that what was, and still is, needed,
for understanding thermodynamic irreversibility, was a method capable of
describing continuous topological evolution. \ \ It was apparent from Cartan's
work \cite{Cartanlecon} that all Hamiltonian processes preserve the Sommerfeld
integrals (closed 1-forms of action), as evolutionary invariants, and could
not describe the dynamics of topological evolution, much less the dynamics of
a radiative transition. \ Clues from prior work had indicated that a
modification of the Hamiltonian method based on Cartan techniques might be
used to explain topological evolution \cite{rmkhamp}. \ Only years later was
this modification recognized to be equivalent to Cartan's magic formula
\cite{Marsden}, where the Lie differential with respect to a direction field,
V, acts on a 1-form of Action, to produce the topological equivalent of the
first law of thermodynamics \cite{vol1}. \ Then it was easy to show
that\ Hamiltonian processes implied that the Heat 1-form, Q, was closed: dQ =
0. \ Recall that the Caratheodory concept of irreversibility was that
Q\symbol{94}dQ $\neq0.$ \ Hence all Hamiltonian processes are
thermodynamically reversible. \ The Cartan topology constructed from the
1-form of Action, was invariant to all Hamiltonian extremal processes. \ The
early objective was determine how to modify the concept of a Hamiltonian
process, and link the idea that topological change was a requirement of
thermodynamic irreversibility \cite{rmkhamp}. \ 

\ Part of the presentation herein will be the demonstration of certain Cartan
techniques that can be used to describe continuous topological evolution and
thermodynamic irreversibility. \ The major objective, however, is to give
examples and methods of construction of closed p-forms, which may serve as the
integrand of period integrals with non-zero values along closed integration
chains which are not boundaries. \ The basic idea stems from the recognition
that the integrands of topological period integrals can be encoded in terms of
homogeneous p-forms of degree zero. \ Homogeneous p-forms of degree not zero
are always exact \cite{PL2}, hence such exact p-forms would yield zero values
for their period integrals along closed integration chains. \ Homogeneous
p-forms of degree zero are independent from "scale changes", not only at a
point, but globally over the homogeneous domain, even though the scale factor
is not a global constant. \ The most common of such objects is to be found in
projective geometry, where the fractional linear, or Moebius transformation,
is used to deduce the important projective invariants. \ All projective
invariants are universally homogeneous functions of cross ratios. \ It is
remarkable that the transition probability of quantum mechanics, according to
Fermi's golden rule, is such a cross ratio invariant.

The concept of "gauge invariance", as introduced by Weyl, was an attempt to
answer the question: Can parallel displacement change the length, or scale, of
a vector. \ Before Weyl, it was recognized that parallel displacement in
Riemannian geometry around closed circuits could change the orientation of a
vector. \ The orientation at the start of the process of parallel transport
around a closed path need not be the same as the orientation of the vector
when it returned to the starting point. \ Orientation changes in the tangent
plane of the starting point were known to be related to curvature, and
orientation changes orthogonal to the tangent plane had been related to the
concepts of torsion. \ Apparently, before Weyl, the idea of a length change
had been ignored in framing the conditions of what was meant by a parallel
displacement in a Riemannian geometry. \ Could the geometric concepts of
metric and connection be reformulated beyond the constraints of Riemannian
geometry to produce scale change and path dependence relative to parallel
transport? \ The details of such reformulations in terms of metrics and
connections appear most cogently in the book by Eddington \cite{EddingtonREL},
and in the concept of Finsler spaces \cite{Antonelli}.

In the language of differential forms, without recourse to geometric
assumptions of metric or connections, the concept of displacement inducing a
change of scale is encoded in terms of the Lie differential with respect to a
direction field, $X=[x^{k}],$ acting on p-forms, $\omega$, to create the same
p-form magnified by a scale factor, $D$. \ A homogeneous differential form
satisfies an equation of the type:%
\begin{equation}
L_{(X)}\omega=D\omega.\label{homopform}%
\end{equation}
Differential forms that satisfy such a formula are said to be homogeneous of
degree $D$. \ The formula is exactly equivalent to Euler's formula for
homogeneous scalar functions, $\Theta(X)$ of homogeneity degree $D.$ \ The
homogeneity index need not be an integer, and the components $x^{k}$ of the
process $X$\ need not be functions of the same (physical) dimension.%
\begin{align}
L_{(X)}\Theta(x^{k})  &  =i(x^{k})d\Theta=\left\langle \partial\Theta/\partial
x^{k}\right\vert \circ\left\vert x^{k}\right\rangle =D\cdot\Theta(x^{k})\\
\Theta(X)  &  =\text{ a zero form.}%
\end{align}

\begin{remark}
Topological evolution can describe changes of scale, without recourse to
specific geometrical constraints.
\end{remark}

The homogeneous formula can be extended to processes, $X$, of arbitrary
direction fields, acting on both pair and impair p-forms. \ Herein, the
concept of relative gauge invariance of functional form is related to
homogeneous functions of degree $D$, and absolute (projective) invariance to
homogeneous functions of degree $D$ equal to zero. \ 

One of the principle results of the first cited article \cite{rmkperiods} was
the presentation and utilization of three period integrals, of dimension 1, 2,
and 3, which have dominant physical significance. \ A period integral is
defined as a closed p-form, $\omega,$ with $d\omega=0,$ integrated over a
(closed)\ cycle of dimension p, $z_{p}$, which is not a boundary. \ In this
article, another 3-dimensional period integral (originally presented in 1977
\cite{rmkames}, \cite{rmkmoffat}) is added to the list. \ The format chosen
will emphasize, for purposes of more rapid comprehension, an electromagnetic
notation and application, but the basic ideas apply to many other areas of
physical speciality, such as hydrodynamics and thermodynamics. \ 

The idea utilizes the topological decomposition of the arbitrary p-form into 3 components:%

\begin{align}
\omega &  =\omega_{no}+\omega_{cl}+\omega_{ex}\\
\omega_{no} &  =\text{Non-closed (Noether potential) component",}\\
\omega_{cl} &  =\text{"Closed but not exact singular component",}\\
\omega_{ex} &  =\text{"Exact component"}%
\end{align}
In another format, the p-form decomposition theorem can be written as%
\begin{align}
\omega^{p} &  =\omega_{no}^{p}+\partial_{z}\omega^{p+1}+d\omega^{p-1},\\
d(d\omega^{p-1}) &  =d(\omega_{ex})=0,\ \ \ \\
d(\partial_{z}\omega^{p+1}) &  =d(\omega_{cl})=0.
\end{align}
This formulation is similar to the Hodge decomposition theorem, but the "cycle
operator", $\partial_{z},$ is not a boundary operator, and definitely is not
the Hodge boundary operator, $\ast d\ast$, which depends upon metric.%

\begin{align}
\partial_{z} &  \neq\partial_{B}\\
\partial_{z} &  \neq\ast d\ast
\end{align}
The cycle operator, $\partial_{z},$ will be defined with examples in the next section.

The decomposition concept goes back at least as far as Hodge-deRham, but the
designation "Noether" component is used herein to tie in with the gr-qc
notation used by Wald \cite{Wald} and others. \ The Wald development utilizes
the fact that "Noether potentials", $\omega_{no},$ of p-forms, upon exterior
differentiation, lead to exact p+1 forms, called Noether currents, whose
closed integrals are evolutionary invariants. \ The integrals of the exact p+1
forms over a domain M are related by Stokes theorem to integrations of
$\omega_{no}$ over the boundary of M. \ The components $\omega_{cl}$ and
$\omega_{ex}$ do not contribute to such integrals over a boundary. \ These
Wald integrals are NOT\ quantized.

Quantized period integrals involve the closed integrals of the closed but not
exact components, $\omega_{cl}.$ \ Such closed but not exact p-forms can be
constructed from a universal algorithm that produces a p-form which is
homogeneous of degree zero relative to its p independent variables and
differentials. \ \ They are quantities defined without use of a metric, and
create closed integrals that are absolute integral invariants relative to any
evolutionary process, $\beta$V, independent from the parametrization
parameter, $\beta.$ \ The notion of a "period" integral is related to the fact
that such structures are singular in the sense that they have fixed points
(singularities of affine transformations) which can be related to physical
concepts of rotation and expansion.

\subsection{Four fundamental topologically quantized period integrals.}

\ The four important topological period integrals (presented here in
electromagnetic format and notation but universal in application) are:

\begin{enumerate}
\item The Flux quantum = $\int_{z1}A_{cl}.$\ \ The integrand\ $A_{cl}$ is a
pair 1-form, and the cycle is a 1-dimensional closed integration chain,
$z_{1}$, in regions where $dA=0.$\ In electromagnetic format the physical unit
of the flux quantum period integral is $h/e$. \ It is important to realize
that the flux quantum is not related to the magnetic flux, nor to the closed
integral of the 2-form, $F=dA.$ \ In hydrodynamics, the flux quantum is
related to the concept of circulation, and is independent from the concept of
vorticity. \ It is important to recognize that the flux quantum occurs only in
domains where the 1-form $A$\ is of Pfaff dimension
%TCIMACRO{\TEXTsymbol{>} }%
%BeginExpansion
$>$
%EndExpansion
1. \ 

\item The Charge quantum = $\int\int_{z2}G_{cl}.$ \ The integrand\ $G_{cl}$ is
an impair 2-form, and the closed cycle is 2-dimensional, $z_{2}$ in domains
where $dG=0$. \ In electromagnetic format the physical unit of the charge
quantum period integral is $e$. \ The fact that the charge quantum is impair
implies that charge is a pseudo-scalar, a fact not in agreement with the
current mainstream convention, but in agreement with the experiments in
crystal physics. \ Recall that integrals of impair forms are not sensitive to orientation.

\item The Topological Torsion or Polarization quantum = $\int\int\int
_{z3}(A\symbol{94}F)_{cl}.$ \ The integrand\ ($A\symbol{94}F)_{cl}$ is a pair
3-form, and the closed cycle is 3-dimensional, $z_{3}$, in a domain where
$d(A\symbol{94}F)=0.$ \ In electromagnetic format the physical unit of the
Topological Torsion quantum period integral is $(h/e)^{2}.$ \ Note that this
physical unit is equal to the spin quantum, $\hbar$, times the Hall
coefficient, $\hbar/e^{2}$. \ Also recall that the non-zero value of
$(A\symbol{94}F)_{cl}$, indicates that the Cartan topology (Chapter
\ref{CTSTRU}) is a disconnected topology, and in a thermodynamic sense implies
that the corresponding thermodynamic system is a nonequilibrium system.

\item The Topological Spin quantum = $\int\int\int_{z3}(A\symbol{94}G)_{cl}$.
\ The integrand\ $(A\symbol{94}G)_{cl}$ is an impair 3-form, and the cycle is
3-dimensional, $z_{3}$ \ in domains where $d(A\symbol{94}G)=0.$ \ In
electromagnetic format the physical unit of the Topological Spin quantum
period integral is $h$. \ The fact that the spin quantum is impair implies
that spin is a pseudo-scalar.
\end{enumerate}

The application of these ideas to EM theory appears in \cite{rmksc3},
\cite{rmktimerev}.

As the integration cycles, $z$, are in domains where the exterior
differentials of the integrands vanish, then the values of the integrals have
rational ratios \cite{deRham} which leads to the idea of topological
"quantization" \ The cycle $z$ wraps around the singular (or fixed) point of
the closed but not exact p-form, which leads to the term "period integral".
\ The integrands for the Flux quantum and Topological Torsion quantum behave
as scalars with respect to transformations of the independent variables in
their arguments. \ Such scalars are, in the language of invariant theory,
called "absolute"\ invariants. \ The closed integrals are sensitive to
orientation. \ 

The Charge quantum and the Topological Spin quantum, are W-densities, and
therefor depend upon the magnitude of determinant of the transformation, but
not upon the sign of the determinant. \ The values of the integrals do not
depend upon the orientation of the domain of integration, nor the fact that
the domain may be non orientable. \ \ Such objects related to the
determinants, in the language of invariant theory are called
"relative"\ invariants \cite{Turnbull}, or pseudo-scalars \cite{PostQuRe}.

The Flux quantum (\symbol{126}Sommerfeld integrals) and the Charge quantum
(\symbol{126}Gauss law) were\ more or less well known in 1977, but the concept
that these period integrals were independent from any metrical constraints was
not so well known. Even now (2006) the fact that these concepts are
independent from metric is not fully appreciated. \ In fact, gravitational
theory emphasizing metric based concepts, has utilized the differential form
argument using closed integrals of 2-form densities to \textit{force} a
relationship between "entropy" and "black hole horizon area" \cite{Jacobsen}.
\ The fact that the closed integrals over W-densities are impair (implying
that the result is a pseudoscalar) is almost completely ignored. \ The idea
that there could exist closed integrals that are period integrals yielding
macroscopic topological quantization at all scales is also almost completely ignored.

In 1977, the third period integral, the Topological Spin quantum, was somewhat
novel, having been discovered\ just a few years before (1969) in a somewhat
different context \cite{rmkintrinsic}. \ \ About the same time \cite{rmkames},
the second 3 dimensional period integral of Topological Torsion was created to
study the topological transition from
\index{Turbulence}
to the streamline state in a fluid. \ It took some 10 to 20 years before it
was appreciated that the nonzero closure of the pair 3-form, $A\symbol{94}F,$
defined domains that could be put into correspondence with thermodynamic
irreversibility. \ In addition, it is only very recently that it has been
appreciated that the 3-form of Topological Torsion has an eigen direction
field that is composed of Spinors, not classical diffeomorphic vectors.
\ Hence in a hydrodynamic context, as a turbulent flow must be irreducibly 4
dimensional, $d(A\symbol{94}F)\neq0$, then the cause of turbulence ultimately
must be traced back to the Spinor content generated by the eigen direction
fields of the 2-form, $dA.$ \ The 3-form $A\symbol{94}F$ is of utmost
importance to (and is nonzero in) the thermodynamic theory of nonequilibrium
systems. \ The use of spinors in macroscopic physics is almost completely
ignored, yet mathematicians have demonstrated that spinors are generators of
minimal surfaces, which are macroscopically observable as wakes and other
modes of propagating tangential discontinuities.

Although each of these period integrals described above\footnote{Yet Torsion
quanta and Spin quanta 3-forms appear sparcely in the literature.} appear to
have application to the microphysical world, an objective of this article is
to emphasize that such macroscopic quantized period integrals also should have
applicability to the cosmological universe. \ After all, period integrals are
topological objects independent from metric constraints of size and shape.
\ The integrands of period integrals are closed p-forms which are homogeneous
of degree zero. \ The p-form, like a cross-ratio in projective geometry, is
independent from metrical scales. \ Size and shape are not important to these
continuous deformation invariants. \ This fact initially posed an ontological
conflict, for experience (or prejudice)\ seems to indicate that "quantum"
features are artifacts of the microphysical world, alone. \ Now it is apparent
that the concept of Spinors is another topological idea based upon the eigen
direction fields of infinitesimal rotations, and does not depend upon scales.
\ Spinors and their importance on macroscopic physical systems has long been ignored.

\begin{remark}
As E. Cartan \cite{Cartanspinors} has demonstrated, Spinors are not vectors
(tensors) with respect to infinitesimal rotations.
\end{remark}

E. J. Post became interested in this predicament, and now champions the idea
that Quantum Mechanics of the microworld should be developed in terms of
metric free ideas \cite{PostTG}. \ On the other hand, the physics of gravity,
constitutive relations, and the synergetic aggregates of the macrophysical
world appear to have geometric, metric-dependent, features. \ Indeed, many of
these geometric features are topological properties, especially when they are
elements of a diffeomorphic equivalence class. \ In order to examine
metric-based topological features, Post recommends the use of general
diffeomorphic invariance principle be used to determine metrical based
topological features. \ That is, the diffeomorphic maps should not be
restricted to some particular geometrical group, such as is presumed in gauge
theories. \ The problem with the use of diffeomorphic maps is that they miss
the discrete symmetry breaking features of handedness of polarization and
to-fro evolution. Diffeomorphic maps imply covariance with respect to both
translations and rotations. \ Spinors are diffeomorphic covariants with
respect to translations, but they are not diffeomorphic covariants with
respect to rotations. \ 

Another method to discover metric independent features is to choose a metric
arbitrarily, and then show (as did Hodge) that certain topological invariants
arise which do not depend upon the choice of metric. \ Such invariants include
those invariants which are "gauge"\ invariant, in the sense that they are
independent from metric based scales. \ At what physical level a metric-based
topology evaporates into a non-metric based topology is still unknown.
\ Conversely, at what level a non-metric based topology condenses or "emerges"
into a metric based topology is intuitively at the level of forming coherent%
\index{Phase (topological)!Coherent Structures}
quantum macro states, such as those that appear in superconductivity, or as
non-dissipative solitons in macro structures. \ It is conjectured that such a
process occurs when the closed, but not exact, homogeneous differential forms
used to construct period integrals become harmonic. \ 

Another suggestive concept that requires investigation is related to how and
if a given metric can undergo topological evolution and change. \ In particular,

\begin{remark}
The signature of a metric may be a process dependent topological feature.
\end{remark}

As mentioned in \cite{vol1} and in more detail in Vol 2, \ \cite{vol2}, the
experimental observations of the features of the nonequilibrium Falaco
Solitons appear to be best represented by a 3D Minkowski metric of signature
\{+,+,-\}, yet the initial state of the fluid, and the ultimate (equilibrium)
state, appear to be Euclidean with a signature \{+,+,+\}. \ If the
observations are correct, the Falaco Solitons \cite{vol2} yield some of the
first experimental results that physics recognizes situations where 3 spatial
dimensions will support a signature which is negative, and non Euclidean. \ \ 

\section{\textbf{Emergent states as coherent topological structures}}

\subsection{1-forms}

Rather than starting with the usual Lagrangian field theory approach
constructed in terms of a Lagrange density N-form and its associated N-1-form
current\footnote{Which are the usual tools of a variational field theory.},
consider those thermodynamic systems that can be encoded in terms of an
exterior differential 1-form of Action, $A$, over a pregeometric (metric not
assigned) variety of dimension N. \ The method is related to the
Cartan-Hilbert invariant integral. \ 

Topological properties of such a 1-form of Action include the Pfaff
topological dimension, which is a statement of the irreducible minimum number
of functions (of the base variables)\ that are required to describe continuous
topological features of the system. \ This minimal number of functions, or
class of a 1-form, can be evaluated by one exterior differentiation, and
subsequent algebraic constructions defined as the Pfaff sequence: \
\begin{equation}
\text{ Pfaff Sequence \ \ }\{A,dA,A\symbol{94}dA,dA\symbol{94}dA...\}.
\end{equation}
\ The number of non-zero entries in the sequence determines the Pfaff
Topological dimension.

As mentioned in the previous section, any 1-form can have three topologically
distinct parts, depending upon the Pfaff topological dimension. \
\begin{align}
A &  =A_{no}+\partial_{z}\omega^{p+1}+d\omega^{p-1},\\
d(d\omega^{p-1}) &  =d(\omega_{ex})=0,\ \ \ \\
d(\partial_{z}\omega^{p+1}) &  =d(\omega_{cl})=0.
\end{align}
If the Pfaff dimension is 1, then only 1-function, say $U(x,y,z,t)$, is
required and $A=dU,\,\ $which is the exact component. \ The "Noether current",
$dA,$ is zero. \ If the Pfaff dimension is 2, then only 2 functions are
required, say $U(x,y,z,t)$ and $V(x,y,z,t).$ A canonical representation is
given by the formulae%
\begin{align}
A  & =UdV\\
dA  & =dU\symbol{94}dV\\
A\symbol{94}dA  & =0.
\end{align}
However there are other possibilities. \ For example, consider the
representation%
\begin{align}
A &  =UdV+\Gamma(U,V)(VdU-UdV)=\\
dA &  =(1-2\Gamma-V\partial\Gamma/\partial V-U\partial F/\partial
U)dU\symbol{94}dV\\
A\symbol{94}dA &  =0.
\end{align}
Only two primitive functions, $U$ and $V$ are required in its construction,
but now the second term has interesting interpretations. \ Orbits of the
second term, can be graphed as rotations if $\Gamma(X,Y)$ is a constant. \ 

\ In general, the second term contributes to the Noether current $dA,$ unless%

\begin{equation}
(Y\partial\Gamma/\partial Y+X\partial F/\partial X)=-2\Gamma,
\end{equation}
which is Euler's equation for homogeneous functions of degree -2. \ In this
special homogenous case, the factor $\Gamma$ becomes an "integrating" factor
for the rotation, such that $dA=0.$ \ In such cases, the rotation is called a
"circulation", and is topologically without limit points, for the "Noether
current" or "vorticity", $dA=0.$ \ It is this construction that defines the
closed but not exact components of the 1-form in terms of a cycle operator
$\partial_{z}$.%

\begin{align}
A_{cl}  & =\partial_{z}\omega^{p+1}=\partial_{z}(dU\symbol{94}dV)\\
& =i([U,V])dU\symbol{94}dV/\lambda\\
& =(UdV-VdU)/\lambda,\\
\lambda & =(aU^{m}+bV^{m})^{(2)/m},\\
\Gamma & =1/\lambda
\end{align}
The coefficients a,b... and the exponent m are constants. \ The function
$\lambda$ is a form of the Holder norm, with a zero set that establishes the
singularities of $A_{cl}.$

Stoke's Law states that%

\begin{align}
\text{ \ for }A  &  =A_{no}+A_{cl}+A_{ex}\\%
%TCIMACRO{\tiint \limits_{M}}%
%BeginExpansion
{\textstyle\iint\limits_{M}}
%EndExpansion
dA  &  =%
%TCIMACRO{\tiint \limits_{M}}%
%BeginExpansion
{\textstyle\iint\limits_{M}}
%EndExpansion
F=%
%TCIMACRO{\tint \limits_{\partial M}}%
%BeginExpansion
{\textstyle\int\limits_{\partial M}}
%EndExpansion
\{A_{no}+A_{cl}+A_{ex}\}=%
%TCIMACRO{\tint \limits_{\partial M}}%
%BeginExpansion
{\textstyle\int\limits_{\partial M}}
%EndExpansion
\{A_{no}\}\\
&  \text{where }\partial M\text{ is a boundary of M.}%
\end{align}
Note that only the Noether term, $A_{no}$, contributes to the integration over
a boundary:%

\begin{equation}%
\begin{array}
[c]{cc}%
\begin{array}
[c]{c}%
\text{For the Noether Component}\\
\text{"The Flux Conservation law"}\\
\text{an absolute evolutionary integral invariant}%
\end{array}
&
%TCIMACRO{\tiint \limits_{M}}%
%BeginExpansion
{\textstyle\iint\limits_{M}}
%EndExpansion
F=%
%TCIMACRO{\tint \limits_{\partial M}}%
%BeginExpansion
{\textstyle\int\limits_{\partial M}}
%EndExpansion
A_{no}\neq0
\end{array}
\end{equation}%
\begin{equation}%
\begin{array}
[c]{cc}%
\begin{array}
[c]{c}%
\text{for the Closed component }\\
\text{Flux quanta balance}%
\end{array}
&
%TCIMACRO{\tiint \limits_{M}}%
%BeginExpansion
{\textstyle\iint\limits_{M}}
%EndExpansion
F=%
%TCIMACRO{\tint \limits_{\partial M}}%
%BeginExpansion
{\textstyle\int\limits_{\partial M}}
%EndExpansion
A_{cl}=0
\end{array}
\end{equation}%
\begin{equation}%
\begin{array}
[c]{cc}%
\text{for the Exact component } &
%TCIMACRO{\tiint \limits_{M}}%
%BeginExpansion
{\textstyle\iint\limits_{M}}
%EndExpansion
F=%
%TCIMACRO{\tint \limits_{\partial M}}%
%BeginExpansion
{\textstyle\int\limits_{\partial M}}
%EndExpansion
A_{ex}=0
\end{array}
\end{equation}
However, integration over a cycle, $z1,$ which is not a boundary and in a
domain where $F=dA=0$, yields%
\begin{equation}%
\begin{array}
[c]{cc}%
\begin{array}
[c]{c}%
\text{for the Closed component}\\
\text{The flux quantum}%
\end{array}
&
%TCIMACRO{\tint \limits_{z1}}%
%BeginExpansion
{\textstyle\int\limits_{z1}}
%EndExpansion
A_{cl}\neq0
\end{array}
\end{equation}%
\begin{equation}%
\begin{array}
[c]{cc}%
\text{for the Exact component} &
%TCIMACRO{\tint \limits_{z1}}%
%BeginExpansion
{\textstyle\int\limits_{z1}}
%EndExpansion
A_{ex}=0
\end{array}
\end{equation}

It is the closed component that yields topological quantization by deRham's
theorems. \ In EM notation the expression for the Bohm-Aharanov flux quantum becomes%

\begin{equation}
\text{Bohm-Aharanov Flux quantum}=%
%TCIMACRO{\tint \limits_{z1}}%
%BeginExpansion
{\textstyle\int\limits_{z1}}
%EndExpansion
A_{cl}\neq0.
\end{equation}
As the integration chain and the integrand are in domains where $F=dA=0,$ the
flux quantum has nothing to do (explicitly) with the classic electromagnetic
flux conservation law, constructed from $%
%TCIMACRO{\tiint }%
%BeginExpansion
{\textstyle\iint}
%EndExpansion
dA(E,B)$, as $dA_{cl}=0.$ \ The flux quantum integral will have values which
are rational multiples of one another, depending upon the cycle, $z1$. \ In
fluid mechanics the closed integral of a closed but not exact velocity field,
such as that encoded by 1-form,
\begin{equation}
A_{cl}(x,y)=\Gamma(ydx-xdy)/(x^{2}+y^{2}),
\end{equation}
defines the circulation integral with value $2\pi\Gamma$, a value that does
not depend upon the 2-form of vorticity. \ Note that if x and y are defined in
terms of a polar coordinate system, $(r,\theta)$, the pullback of
$A_{cl}(x,y)$ becomes%

\begin{equation}
A_{cl}(r,\theta)=\Gamma_{0}d(\theta)\Leftarrow\Gamma_{0}(ydx-xdy)/(x^{2}%
+y^{2})=A_{cl}(x,y).
\end{equation}
It would appear the 1-form $A_{cl}(x,y)$ when pulled back to the space of
variables $\{r,theta\}$ is an exact differential, $d(\theta).$ \ The notation
is decieving, but it must be remembered that $\theta$ as used above is a
cyclic variable; \ the coordinate mapping fails at r = 0. \ The excluded
point, $r=0,$ represents a topological defect, a hole in the Cartesian fabric
of 2 dimensions.

Many other examples of constructing deRham period integrals in terms of
homogeneous p-forms can be found in chapter 8 of \cite{vol1}.

\subsection{2-form densities}

As the Wald description of blackhole entropy has a realization in terms of
"Noether" currents (3-forms), it is of some interest to formulate the concept
of impair 2-form densities (the Noether "potentials). \ This will be done
first in terms of EM notation .

\subsubsection{EM notation \ \ \ \ }

The story for 2-form densities is comparable to the story for\ 1-forms given
above. \ Consider the impair 2-form density $G$ in EM notation, (or $Q$ in GR
notation):%
\begin{align}
\text{"Noether potential" component, }G_{no} &  :dG_{no}\neq0\\
\text{Closed but not Exact singular component, }G_{cl}. &  :dG_{cl}=0\\
\text{Closed and Exact component, }G_{ex} &  :dG_{ex}=0.
\end{align}

Stoke's Law states that%
\begin{align}
\text{ \ \ for }G  &  =G_{no}+G_{c\text{ }l}+G_{ex}\\%
%TCIMACRO{\tiiint \limits_{M}}%
%BeginExpansion
{\textstyle\iiint\limits_{M}}
%EndExpansion
dG  &  =%
%TCIMACRO{\tiiint \limits_{M}}%
%BeginExpansion
{\textstyle\iiint\limits_{M}}
%EndExpansion
J=%
%TCIMACRO{\tiint \limits_{\partial M}}%
%BeginExpansion
{\textstyle\iint\limits_{\partial M}}
%EndExpansion
\{G_{no}+G_{cl}+G_{ex}\}=%
%TCIMACRO{\tiint \limits_{\partial M}}%
%BeginExpansion
{\textstyle\iint\limits_{\partial M}}
%EndExpansion
\{G_{no}\}\\
&  \text{where }\partial M\text{ is a boundary of M.}%
\end{align}

Note that only the Noether term, $G_{no}$, contributes to the integration over
a boundary:%
\begin{equation}%
\begin{array}
[c]{cc}%
\begin{array}
[c]{c}%
\text{for the Noether component}\\
\text{"The Charge Conservation law" }\\
\text{an absolute evolutionary integral invariant}%
\end{array}
&
%TCIMACRO{\tiiint \limits_{M}}%
%BeginExpansion
{\textstyle\iiint\limits_{M}}
%EndExpansion
J\Rightarrow%
%TCIMACRO{\tiint \limits_{\partial M}}%
%BeginExpansion
{\textstyle\iint\limits_{\partial M}}
%EndExpansion
G_{no}\neq0
\end{array}
\end{equation}%
\begin{equation}%
\begin{array}
[c]{cc}%
\begin{array}
[c]{c}%
\text{For the closed component:}\\
\text{Equal and opposite charge pairs cancel}\\
\text{charge neutrality as an impair effect}%
\end{array}
& \Rightarrow%
%TCIMACRO{\tiint \limits_{\partial M}}%
%BeginExpansion
{\textstyle\iint\limits_{\partial M}}
%EndExpansion
G_{cl}=0
\end{array}
\end{equation}%
\begin{equation}%
\begin{array}
[c]{cc}%
\text{for the Exact component } & \Rightarrow%
%TCIMACRO{\tiint \limits_{\partial M}}%
%BeginExpansion
{\textstyle\iint\limits_{\partial M}}
%EndExpansion
G_{ex}=0
\end{array}
\end{equation}
However, integration over a cycle, $z2,$ which is not a boundary and in a
domain where $J=dG=0$, yields%
\begin{equation}%
\begin{array}
[c]{cc}%
\begin{array}
[c]{c}%
\text{for the Closed component}\\
\text{The Charge quantum}%
\end{array}
&
%TCIMACRO{\tiint \limits_{z2}}%
%BeginExpansion
{\textstyle\iint\limits_{z2}}
%EndExpansion
G_{cl}\neq0
\end{array}
\end{equation}

\begin{equation}%
\begin{array}
[c]{cc}%
\text{for the Exact component} &
%TCIMACRO{\tiint \limits_{z2}}%
%BeginExpansion
{\textstyle\iint\limits_{z2}}
%EndExpansion
G_{ex}=0
\end{array}
.
\end{equation}

It is the singular closed but not exact component that yields topological
quantization by deRham's theorems. \ In EM notation the expression for the
Charge quantum becomes an integration over a cycle, not a boundary,%

\begin{equation}
\text{Charge quantum}=%
%TCIMACRO{\tiint \limits_{z2}}%
%BeginExpansion
{\textstyle\iint\limits_{z2}}
%EndExpansion
G_{cl}\neq0.\label{Qq}%
\end{equation}
A construction for representing a closed but not exact component of a 2-form,
$G_{cl},$ follows the same procedure given for the closed but not exact
1-form. \ The 2-form is defined in terms of a cycle operator $\partial_{z}$
acting on a 3-form. \ The 3-form is constructed as the monomial differential
volume element of three arbitrary independent functions, $\{U,V,W\}$ (over the
N=4 base variables). \ The cycle operator is defined in terms of the 2-form
"current" multiplied by an integrating factor, $1/\lambda(U,V,W)$ such that
$dG_{cl}=0.$ \
\begin{align}
G_{cl}  & =\partial_{z}\omega^{p+1}=\partial_{z}(dU\symbol{94}dV\symbol{94}%
dW)\\
& =i([U,V,W])dU\symbol{94}dV\symbol{94}dW/\lambda\\
& =(UdV\symbol{94}dW-VdU\symbol{94}dW+WdU\symbol{94}dV)/\lambda,\\
\lambda & =(aU^{m}+bV^{m}+eW^{m})^{(3/m)},\\
\Gamma & =1/\lambda,\ \ \ \ dG_{cl}=0.
\end{align}
The closed non-exact component of the 2-form, $G_{cl}$, is homogeneous of
degree zero in terms of its functions. \ The choice of integrating factor
given above is based on an extension of the Holder norm. \ The homogeneous
2-form, $G_{cl}$, has many representations in terms of the arbitrary constants
(signature) $\{a,b,c\}$ and the exponent $m$. \ Note that the choice of the
cubic format, $m=3,$ yields a simple algebra.

\subsection{3-form Currents}

The construction for the closed but not exact component of a 3-form follows
the procedure given for the 1-form. \
\begin{align}
J_{cl}  & =\partial_{z}\omega^{p+1}=\partial_{z}(dU\symbol{94}dV\symbol{94}%
dW\symbol{94}dS)\\
& =i([U,V,W,S])dU\symbol{94}dV\symbol{94}dW\symbol{94}dS/\lambda\\
& =(UdV\symbol{94}dW\symbol{94}dS-VdU\symbol{94}dW\symbol{94}dS+WdU\symbol{94}%
dV\symbol{94}dS-SdU\symbol{94}dV\symbol{94}dW)/\lambda,\\
\lambda & =(aU^{m}+bV^{m}+eW^{m}+fS^{m})^{(4/m)},\\
\Gamma & =1/\lambda,\ \ \ \ \ dJ_{cl}=0.
\end{align}
The closed non-exact component of the 3-form is homogeneous of degree zero in
terms of its functions.

The results constructed above for 1, 2, and 3-forms can be generalized as a theorem.

\begin{theorem}
On a pregeometric variety of N independent base variables, a projective
differential volume element of M independent functions, $d(Vol)=dV^{1}%
\symbol{94}...dV^{M},$ can always be associated with an M-1 current,
$J=i([V^{1},..,V^{M})d(Vol)$ that admits an integrating factor of the form
$1/\lambda$ where $\lambda=((a1(V^{1})^{m}...+am(V^{M}))^{(M/m)}$ such that
$d(J/\lambda)=0,$ and the renormalized current is homogeneous of degree zero.
\ It is thereby possible to construct an infinite number of conservation laws
on an N volume.
\end{theorem}

\begin{remark}
I\ was led to this theorem from a study of singularities presented in Chapter
2 of Sewell \cite{Sewell}, especially the problem 2.3.3 on page 108. \ The
idea of a homogeneity "integrating factor" can also be accomplished in terms
of the Buckingham Pi product, a format which is utilized in the same problem.
\end{remark}

\subsubsection{GR QC notation}

In GR applications \cite{Jacobsen}, the notation changes but the game is the
same. \ \ Merely substitute the symbol $Q$ in the 2-form expressions above,
such that $Q=Q_{no}+Q_{cl}+Q_{ex}$. \ Then the "Noether Current" (as used by
Wald and Jacobsen) is defined \ as $J=dQ=dQ_{no},$ and does not depend upon
either the closed or exact components of $Q.$ \ Wald uses the idea that $Q$ is
the "Noether potential" for which "entropy" is defined as
\begin{equation}
"\text{Entropy" }S=2\pi%
%TCIMACRO{\toint }%
%BeginExpansion
{\textstyle\oint}
%EndExpansion
Q,
\end{equation}
which unfortunately gives the (incorrect) impression that this formula is
somehow related to a 1 dimensional integral of some 1-form. \ Indeed, it
appears that the expression for the equilibrium thermodynamic system described
by the formula,
\begin{equation}
PdV+dU-TdS=0,
\end{equation}
motivated the early conjectures about "Black Hole Entropy". \ This expression
of the first law in isolated equilibrium systems is indeed an exterior
differential system based upon a 1-form. \ However, the method employed by
Wald and Jacobsen is not related to such a 1-form, but instead is related to
the evaluation of a 2-form over a boundary which is an area. \ A somewhat more
precise notation for the Wald formula would be written as:%

\begin{align}
\text{"Entropy"\ }S  &  =2\pi%
%TCIMACRO{\tiint \limits_{\partial M}}%
%BeginExpansion
{\textstyle\iint\limits_{\partial M}}
%EndExpansion
Q=2\pi%
%TCIMACRO{\tiint \limits_{\partial M}}%
%BeginExpansion
{\textstyle\iint\limits_{\partial M}}
%EndExpansion
\{Q_{no}+Q_{cl}+Q_{ex}\}\\
&  =2\pi%
%TCIMACRO{\tiint \limits_{\partial M}}%
%BeginExpansion
{\textstyle\iint\limits_{\partial M}}
%EndExpansion
\{Q_{no}\}+0+0\\
&  =%
%TCIMACRO{\tiiint \limits_{M}}%
%BeginExpansion
{\textstyle\iiint\limits_{M}}
%EndExpansion
dQ_{no}=%
%TCIMACRO{\tiiint \limits_{M}}%
%BeginExpansion
{\textstyle\iiint\limits_{M}}
%EndExpansion
J\neq0,\text{ \ }\\
J  &  =dQ_{no}\ \ \ \text{defined as the "Noether" current.}%
\end{align}
It is not at all clear that this formulation has anything to do with
Thermodynamic entropy. \ Note that by merely changing the letters, the
formalism is exactly that given above relating to the Charge-Current 4 vector
of electromagnetism and the conservation of charge-current in EM theory. \ %

\begin{equation}%
%TCIMACRO{\tiint \limits_{\partial M}}%
%BeginExpansion
{\textstyle\iint\limits_{\partial M}}
%EndExpansion
G_{no}=%
%TCIMACRO{\tiiint \limits_{M}}%
%BeginExpansion
{\textstyle\iiint\limits_{M}}
%EndExpansion
dG_{no}=%
%TCIMACRO{\tiiint \limits_{M}}%
%BeginExpansion
{\textstyle\iiint\limits_{M}}
%EndExpansion
J\neq0
\end{equation}
From the topological perspective, changing the symbols does not change the
universality of the ideas. \ Should I then believe that the Entropy - Area
formula is nothing more than using different symbols, but is equivalent to
Gauss' law relating a surface area integration of the D field on a boundary to
the integral of the charge density in the bounded volume in EM theory? \ Is
the integral of G(D,H) over a bounding area somehow related to entropy? \ What
has charge to do with Entropy? \ Do not these questions leave the Bekenstein -
Hawking concept of black hole entropy, and especially Wald's formulation
somewhat suspect, and perhaps the result of speculative wishful thinking.
\ The Wald integrals have nothing to do with topological quantization.

In my opinion, the (somewhat suspect) Wald formulation\ does open Pandora's
box. \ What about the possible quantum features? \ Could it be that there
exist cosmological quanta associated with period integrals of a closed but
non-exact, $Q_{cl}$? \ These possibilities will be discussed below. \ First it
is necessary to discuss the difference between pair and impair differential
forms, and their relationships to Lagrangian N-form densities. \ 

\subsection{Lagrangian pair and impair N-forms}

Consider maps defining the range of vector arrays with coefficients, $V^{m},$
as functions of the domain of independent variables $x^{k}$ :%

\begin{align}
\phi &  :x^{k}\Rightarrow V^{m}=V^{m}(x^{k})\\
d\phi &  :dx^{k}\Rightarrow dV^{m}=\{\partial V^{m}(x^{k})/\partial
x^{n}\}dx^{n}.
\end{align}
These maps $\phi$\ need \textit{not} be diffeomorphisms. \ The function
$\Delta$ is defined as the determinant of the mapping Jacobian matrix, \
\begin{equation}
\Delta(x^{m})=\det[\partial V^{m}(x^{k})/\partial x^{n}],
\end{equation}
which is not zero, if the map is a diffeomorphism. \ Another construction
defines the sign of the determinant as%

\begin{equation}
\left\vert \Delta\right\vert /\Delta=sign[\partial V^{m}(x^{k})/\partial
x^{n}],
\end{equation}

\subsubsection{Pair and Impair}

Consider various field functions defined on the range variables, $V^{m}$, and
collectively named $\varphi(V^{m}).$ \ Next, consider a special function
L\ (the Lagrange function) of these variables, denoted by the symbol
L$(\varphi(V^{m}))=\rho(V^{m}).$ \ 

There now are two possibilities: \ 

\begin{enumerate}
\item Construct the Pair N-form on the range $dV^{m}:\{\rho\}dV^{1}%
\symbol{94}dV^{2}..\symbol{94}dV^{N}$\bigskip

\item or the Impair N-form on the range $dV^{m}:\{\rho\cdot(\left\vert
\Delta\right\vert /\Delta\}dV^{1}\symbol{94}dV^{2}..\symbol{94}dV^{N}$
\end{enumerate}

Use functional substitution defined by the map $\phi$ and its differentials to
evaluate the pullbacks of both the pair and impair p-forms:
\begin{equation}
\text{Using }dV^{1}\symbol{94}dV^{2}..\symbol{94}dV^{N}\Rightarrow\Delta
(x^{m})dx^{1}\symbol{94}dx^{2}..\symbol{94}dx^{N},
\end{equation}

\begin{align}
\text{Pair N-form }\{\rho(x^{k})\Delta(x)\}dx^{1}\symbol{94}dx^{2}%
..\symbol{94}dx^{N}  &  \Leftarrow\{\rho(V)\}dV^{1}\symbol{94}dV^{2}%
..\symbol{94}dV^{N},\\
\text{A "scalar\ }\Delta\text{-density"}  &  \text{:}\rho(x^{k})\Delta
(x)\Leftarrow\rho(V).
\end{align}%
\begin{align}
\text{Impair N-form \ }\{\rho(x^{k})\left\vert \Delta(x)\right\vert
\}dx^{1}\symbol{94}dx^{2}..\symbol{94}dx^{N}  &  \Leftarrow\{\rho
(V)\cdot(\left\vert \Delta\right\vert /\Delta)\}dV^{1}\symbol{94}%
dV^{2}..\symbol{94}dV^{N},\\
\text{A\ "pseudoscalar }\Delta\text{-density"}  &  \text{:}\{\rho
(x^{k})\left\vert \Delta(x)\right\vert \}\Leftarrow\rho(V)(\left\vert
\Delta\right\vert /\Delta).
\end{align}
The important thing to remember is that the integrals of Pair p-forms depends
upon the sign of the orientation of the integrand, and the integrals of Impair
p-forms do not depend upon sign of the orientation of the integrand. \ %

\begin{align}
\text{Scalar-densities }  &  \text{:}\rho(x^{k})\cdot\Delta(x)\\
\text{PseudoScalar-densities}  &  \text{: }\rho(x^{k})\cdot\left\vert
\Delta(x)\right\vert
\end{align}

The Lagrangian N-form can have two representations (or sometimes the complex
N-form that consists of both the Pair and the Impair structure). \ One
representation recognizes that orientation is important, so that the
Lagrangian is written as a pair N-form on the range space:%

\begin{equation}
\text{Pair Lagrangian N-form}=\{\text{L}(\varphi(V^{k}))\}dV^{1}%
\symbol{94}dV^{2}..\symbol{94}dV^{N}\}.
\end{equation}
Next construct the Lie differential of the N-form as%

\begin{equation}
L_{(V^{m})}\{\text{L}(\varphi)\}dV^{1}\symbol{94}dV^{2}..\symbol{94}%
dV^{N}\}=d\{\text{L}(\varphi)i(V^{m})dV^{1}\symbol{94}dV^{2}..\symbol{94}%
dV^{N}\}.\label{dfnform}%
\end{equation}
Hence there exists an N-1 form Current of the format:%

\begin{equation}
J(V)=\text{L}(\varphi)\{i(V^{m})dV^{1}\symbol{94}dV^{2}..\symbol{94}dV^{N}\}
\end{equation}
that pulls back to the domain space as%

\begin{align}
\text{Pair }J(x)  &  =\text{L}(\psi(x))\Delta(x)\{i(V^{m})dx^{1}%
\symbol{94}dx^{2}..\symbol{94}dx^{N}\},\\
\psi(x)  &  =\varphi(V(x)).
\end{align}
The formula for the Impair pullback is (the only difference is the use of the
absolute magnitude of the determinant):%

\begin{align}
\text{Impair }J(x)  &  =\text{L}(\psi(x))\left\vert \Delta(x)\right\vert
\{i(V^{m})dx^{1}\symbol{94}dx^{2}..\symbol{94}dx^{N}\},\\
\psi(x)  &  =\varphi(V(x)).
\end{align}

\begin{remark}
The integrals of pair forms depend upon the choice of orientation of the
integration chain. \ 

The integrals of impair forms do not depend upon an orientation of the
integration chain.
\end{remark}

\subsection{Thermodynamic Quantized Currents (3-forms)}

\subsubsection{The exterior differential form method}

The idea is to use topological thermodynamics (where physical systems are
encoded in terms of various 1-forms and 2-forms and p-forms), and exterior
calculus of Cartan to algebraically deduce 3-form currents with their Noether
components and their closed components. \ The method is to be compared with
the ubiquitous, but topologically awkward, Lagrangian approach that is based
upon a starting point of N-form densities, and their associated Noether
currents, but leaves undetermined the closed and the exact components of these
currents. \ Note that in physical thermodynamic systems both species of pair
and impair p-forms are useful. \ Pair forms are related to "intensities" such
as pressure and temperature, while impair p-forms are related to "additive
quantities, or source excitations" such as volumes and entropy. \ 

The most familiar examples of the thermodynamic method are exhibited by the
topological features \ of electromagnetism. \ In EM theory, the 2-form of
intensities, $F(E,B)=dA,$ is a pair 2-form (which is exact), and the 2-form of
excitations, $G(D,H),$ is impair. \ These facts have been experimentally
verified from studies of the behavior of electromagnetic signals in
crystalline media with and without a center of symmetry \cite{PostTG}. \ The
symbols of EM theory will be used in this section, as the notation is more
familiar to most (physicist)\ readers. \ It does not mean that the ideas are
restricted to an EM\ interpretation. \ Thermodynamics is universal to all
physical systems.

An example of a Pair 4-form algebraically can be constructed from the 1-form,
$A,$ and its 2-form, $F=dA,$ such that the exterior product of $A\symbol{94}F$
generates a Pair 3-form "current", $J_{pair}=H=A\symbol{94}F$. \ This 3-form
is a Pair 3-form as $F$\ is a Pair 2-form, and $A$ is a pair 1-form. \ This
form I\ have called the 3-form of Topological Torsion. \ It has the usual
3-part decomposition in terms of Noether, closed, and exact components. \
\begin{align}
\text{Pair } &  :\text{\ Topological Torsion 3-form}\\
J_{pair} &  \Rightarrow H=A\symbol{94}F\text{ \ }\\
&  =H_{no}+H_{cl}+H_{ex}%
\end{align}
The 3-form is a current that can be explicitly determined from the formula for
the Topological Torsion Vector, $\mathbf{T}_{4}$, such that
\begin{equation}
i(\mathbf{T}_{4})d(Vol)=A\symbol{94}F.
\end{equation}
It is remarkable that the 4 components (relative to x,y,z,t) of this vector
can be evaluated in terms of the functions (and their partial differentials)
that define the 1-form of Action, $A$,
\begin{equation}
\mathbf{T}_{4}=[\mathbf{E\times A}+\mathbf{B}\phi,~\mathbf{A\cdot B]}%
\end{equation}

\subsubsection{The Second Poincare Invariant (a Pair 4-form)}

The exterior differential of the Topological Torsion\footnote{Sometimes
refered to as the Helicity 3-form} current 3-form leads to the Topological
Parity 4-form, $K=F\symbol{94}F$. \ The closed integrals of $F\symbol{94}F$
define the Second Poincare Invariant.
\begin{align}
\text{Pair }  &  :\text{Topological Parity 4-form}\\
dH  &  =dH_{no}=d(A\symbol{94}F)=F\symbol{94}F=K\\
K  &  =F\symbol{94}F=2(\mathbf{E\cdot B})dx\symbol{94}dy\symbol{94}%
dz\symbol{94}dt\\%
%TCIMACRO{\tiiiint \limits_{closed}}%
%BeginExpansion
{\textstyle\iiiint\limits_{closed}}
%EndExpansion
F\symbol{94}F  &  =\text{ Second Poincare Invariant.}%
\end{align}
When evaluated in EM symbols, it is apparent that the ($\Delta)$ density
coefficient of $F\symbol{94}F$ is\ $2(\mathbf{E\cdot B}).$ \ 

The second Poincare invariant, indeed, is an evolutionary invariant, for the
continuous topological evolution generated by the Lie differential (with
respect to \textit{any} evolutionary direction field, $\beta V^{k}$) acting on
the closed integrals of $F\symbol{94}F$ is zero. \ That is, continuous
topological evolution produces no change in the closed integrals of
$F\symbol{94}F$:%

\begin{align}
L_{(\beta V^{k})}%
%TCIMACRO{\tiiiint \limits_{closed}}%
%BeginExpansion
{\textstyle\iiiint\limits_{closed}}
%EndExpansion
F\symbol{94}F  &  =%
%TCIMACRO{\tiiiint \limits_{closed}}%
%BeginExpansion
{\textstyle\iiiint\limits_{closed}}
%EndExpansion
\{i(\beta V^{k})d(F\symbol{94}F)+d(i(\beta V^{k})F\symbol{94}F)\}\\
&  =%
%TCIMACRO{\tiiiint \limits_{closed}}%
%BeginExpansion
{\textstyle\iiiint\limits_{closed}}
%EndExpansion
\{0+d(i(\beta V^{k})F\symbol{94}F)\}=0.
\end{align}
Note that the evolutionary invariance of the closed integral is valid
independent from the parameterization factor, $\beta(x,y,z,t),$ of the
direction field, $V^{k}.$

It is further remarkable that evolution of the Cartan topology (generated\ by
the 1-form of Action) in the direction\ of the topological Torsion vector,
$\mathbf{T}_{4},$ is thermodynamically irreversible when $F\symbol{94}F$ is
not zero, for%

\begin{align}
L_{(\mathbf{T}_{4}^{k})}A &  =(\mathbf{E\cdot B})A=Q\\
L_{(\mathbf{T}_{4}^{k})}dA &  =d(\mathbf{E\cdot B})\symbol{94}%
A+(\mathbf{E\cdot B})dA=dQ,\\
Q\symbol{94}dQ &  =(\mathbf{E\cdot B})^{2}(A\symbol{94}dA)\neq0
\end{align}
The fact that $Q\symbol{94}dQ$ is NOT zero implies that $Q$ does not admit an
integrating factor, which is the classical idea \cite{Morse} that the process
that generated $Q$ is thermodynamically irreversible. \ The fact that
$\mathbf{E\cdot B}$ cannot be zero implies that the Pfaff topological
dimension of the 1-form of Action, $A$, must be 4.

The topological evolution of the volume element with respect to the
irreversible process represented by $\mathbf{T}_{4}$ is given by the
expression:%
\begin{equation}
L_{(\mathbf{T}_{4}^{k})}d(Vol)=2(\mathbf{E\cdot B})d(Vol).
\end{equation}
The dissipative irreversible evolution of the volume element can be positive
or negative, representing an expansion or contraction of space time, depending
upon the sign of the dissipation coefficient, $(\mathbf{E\cdot B}). $ \ 

\begin{remark}
The expanding universe is an artifact of thermodynamic irreversibility.
\end{remark}

\subsubsection{The First Poincare Invariant (an Impair 4-form)}

An example of a Impair 4-form can be given by the expression related to the
First Poincare invariant, a portion of which is often used to define the
electromagnetic impair Lagrange density for the electromagnetic field. \ The
impair 4-form also has a Current that can be written in the format of the
impair 3-form $A\symbol{94}G$, herein called "Topological Spin". \ \ This
3-form I\ have called the 3-form of Topological Spin. \ It has the usual
3-part decomposition in terms of Noether, closed, and exact components. \
\begin{align}
\text{Impair } &  :\text{\ Topological Spin 3-form}\\
J_{impair} &  \Rightarrow S_{impair}=A\symbol{94}G\text{ \ }\\
&  =S_{no}+S_{cl}+S_{ex}%
\end{align}

The exterior differential of the Topological Spin current 3-form leads to the
Lagrange density 4-form, \textbf{L}. \ The closed integrals of \textbf{L}
define the Second Poincare Invariant.
\begin{align}
\text{Impair } &  :\text{Lagrange density 4-form}\\
dS &  =dS_{no}=d(A\symbol{94}G)=F\symbol{94}G-A\symbol{94}J=\text{\textbf{L}%
}\\
\text{\textbf{L}} &  \mathbf{=}F\symbol{94}G-A\symbol{94}J\\
&  =\{(\mathbf{B\cdot H-D\cdot E})-(\mathbf{A\cdot J}-\rho\phi)\}dx\symbol{94}%
dy\symbol{94}dz\symbol{94}dt,\\%
%TCIMACRO{\tiiiint \limits_{closed}}%
%BeginExpansion
{\textstyle\iiiint\limits_{closed}}
%EndExpansion
F\symbol{94}G-A\symbol{94}J &  =\text{ First Poincare Invariant.}%
\end{align}

The topological 4-form (deducible from the topological Spin 3-form,
$A\symbol{94}G)$\ was "discovered" in 1974 \cite{rmkintrinsic}.

The second Poincare invariant, indeed, is an evolutionary invariant, for the
continuous topological evolution generated by the Lie differential (with
respect to \textit{any} evolutionary direction field, $\beta V^{k}$) acting on
the closed integrals of $F\symbol{94}G-A\symbol{94}J=d(A\symbol{94}G)$ is
zero. \ That is, continuous topological evolution produces no change in the
closed integrals of $d(A\symbol{94}G)$:%

\begin{align}
L_{(\beta V^{k})}%
%TCIMACRO{\tiiiint \limits_{closed}}%
%BeginExpansion
{\textstyle\iiiint\limits_{closed}}
%EndExpansion
d(A\symbol{94}G)  &  =%
%TCIMACRO{\tiiiint \limits_{closed}}%
%BeginExpansion
{\textstyle\iiiint\limits_{closed}}
%EndExpansion
\{i(\beta V^{k})dd(A\symbol{94}G)+d(i(\beta V^{k})d(A\symbol{94}G)\}\\
&  =%
%TCIMACRO{\tiiiint \limits_{closed}}%
%BeginExpansion
{\textstyle\iiiint\limits_{closed}}
%EndExpansion
\{0+d(i(\beta V^{k})d(A\symbol{94}G)\}=0.
\end{align}
Note that the evolutionary invariance of the closed integral of the first
Poincare 4-form is valid independent from the parameterization factor,
$\beta(x,y,z,t),$ of the direction field, $V^{k}.$

\subsubsection{Period integrals of 3-forms}

Period integrals are integrals over closed integration chains that are not
boundaries of closed but not exact forms. \ Each 3-form is a current of the
format$:$%
\begin{equation}
J_{pair}=J_{no}+J_{cl}+J_{ex}.
\end{equation}
Period integrals have integrands which are closed but not exact. \ Hence the
domains of interest are where the Noether component is zero, and the exact
component is of no consequence to the value of the integral.%

\begin{equation}
\text{3-form Period Integral}=%
%TCIMACRO{\tiiint \limits_{z3}}%
%BeginExpansion
{\textstyle\iiint\limits_{z3}}
%EndExpansion
J_{cl}.
\end{equation}
By deRham's theorems, the value of the integral is an integer times a constant
depending upon the cycle z3. \ 

There are two types of period integrals, depending upon whether the integrand
is Pair or Impair. \ The Pair 3-form, $A\symbol{94}F$, of topological Torsion
has possible periods in domains where the second Poincare invariant vanishes,
$F\symbol{94}F=0.$ \ In such domains, the Pfaff topological dimension of $A$
must be $<4.$ \ As $A\symbol{94}F$ vanishes in domains of Pfaff topological
dimension $<3$, it follows that period integrals of Torsion must exist in
domains of Pfaff dimension 3. \ If continuous topological evolution causes a
domain of Pfaff dimension 3 to emerge (like a condensation)\ from the physical
vacuum of Pfaff dimension 4, then such domains could be topologically
quantized. \ The period integral 3-form\ $%
%TCIMACRO{\tiiint \limits_{z3}}%
%BeginExpansion
{\textstyle\iiint\limits_{z3}}
%EndExpansion
J_{cl}$ would have values n times a constant representing physical units. \ In
EM theory the physical units of $A\symbol{94}F$ are $(\hbar/e)^{2}%
=Z_{Hall}\cdot\hbar.$ \ Hence the periods for $J_{cl}=H_{cl}$ would be of the form,%

\begin{equation}
\text{Topological Torsion Quanta: }%
%TCIMACRO{\tiiint \limits_{z3}}%
%BeginExpansion
{\textstyle\iiint\limits_{z3}}
%EndExpansion
J_{cl}\Rightarrow%
%TCIMACRO{\tiiint \limits_{z3}}%
%BeginExpansion
{\textstyle\iiint\limits_{z3}}
%EndExpansion
H_{cl}=\pm\ n~(Z_{Hall}\cdot\hbar).
\end{equation}
The periods are sensitive to the orientation of the integration chain, hence
they have plus and minus values. \ For the 3-form of topological torsion to be
closed it is necessary that $F\symbol{94}F=0=2(\mathbf{E\circ B}%
)dx\symbol{94}dy\symbol{94}dz\symbol{94}dt$. \ The constraint means the second
Poincare invariant must be zero. \ The macroscopic domains that satisfy the
conditions of a period integral, are defined as topological Torsion quanta.

Similarly, when the first Poincare Invariant vanishes, that is in domains
where the exterior differential of the impair 3-form, $A\symbol{94}G$
vanishes, closed but not exact components of $A\symbol{94}G$ can have
"quantized" period integrals in the sense of deRham. \ That is, the integrals
of the closed but not exact integrands, relative to closed integration chains
which are cycles and not boundaries, can have values which are rational
numbers times a constant. \ In EM theory, the physical units of $A\symbol{94}%
G$ are $\hbar,$such that the periods of the 3-form $A\symbol{94}G$ become:%
\begin{equation}
\text{Topological Spin Quanta: }%
%TCIMACRO{\tiiint \limits_{z3}}%
%BeginExpansion
{\textstyle\iiint\limits_{z3}}
%EndExpansion
J_{cl}\Rightarrow%
%TCIMACRO{\tiiint \limits_{z3}}%
%BeginExpansion
{\textstyle\iiint\limits_{z3}}
%EndExpansion
S_{cl}=n~(\hbar).
\end{equation}
Such objects are defined as Topological Spin quanta. \ The macroscopic domains
of topological Spin quanta admit several possibilities, as the necessary
algebraic constraint of closure, $d(A\symbol{94}G)\Rightarrow0:$
\begin{equation}
\{(\mathbf{B\cdot H}-\mathbf{D\cdot E})-(\mathbf{A\cdot J}-\rho\phi
)\}\Rightarrow0,
\end{equation}
can be satisfied in several ways. \ The Spin quanta are not sensitive to the
orientation of the integration domain.

\section{Conclusions}

In Part I the idea was to describe a topological cosmology in terms of an open
thermodynamic system. \ Such systems can be encoded by means out a 1-form of
action, A, of Pfaff topological dimension 4. \ Such systems support continuous
topological evolution to states of the lower Pfaff dimension, and such states
may be viewed as topological defects in the Open system environment. \ These
states emerge from the open system forming topologically coherent structures
of Pfaff dimension 3, which are not in thermodynamic equilibrium in the sense
that they exchange radiation with their environment. \ They represent stars
and galaxies which have "condensed" out of the background cosmology.

These Pfaff dimension 3 states can continue to evolve to thermodynamic states
of lower Pfaff dimension that represent isolated or equilibrium systems.
\ Rotational spirals to a limit cycle are typical artifacts of such processes.
\ It is remarkable that one of the possible evolutionary processes supported
\ by thermodynamic systems of Pfaff dimension 3 is a Hamiltonian process,
where the topology remains constant for appreciable lifetimes (mod topological fluctuations).

In Part II it was noted that during processes of continuous topological
evolution it is possible that topologically coherent macroscopic (at all
scales) states will emerge which have closed integrals that are proportional
to the integers. These are the states of a quantum cosmology, and they are
seen at all scales. \ The key feature is that these states are represented by
p-forms which are homogeneous of degree zero (self-similarity). \ Mathematical
examples of such singularity defect structures appear at Pfaff topological
dimensions of 1, 2, and 3. \ These topologically quantized states - the flux
quantum, the charge quantum, the torsion quantum and the spin quantum - are
not dependent upon metric issues. \ 

An open question is the description of a black hole in terms of thermodynamic
states. \ It must be a non equilbrium state for it is accepting radiation and
matter from its enviroment. \ For excited states (with a black hole
temperature) it could be thermodynamically closed in the sense that it
exchanges radiation with its environment; \ but mass is not exchanged, only
absorbed. \ It is conjectured that the one way process for mass is to be
associated with fact that 3-form of topological Spin is not sensitive to
orientation. \ On the otherhand, radiation and the 3-form of topological
Torsion is orientation (polarization)\ sensitive. \ These issues are under investigation.

The bottom line is that quantum cosmology is a topological issue not a
geometrical issue.

\end{document}